\documentclass[twocolumn]{autart}
\usepackage[a4paper,margin={1.55 cm,1.55 cm,1.55 cm,1.55 cm}]{geometry}  % Enable this line and disable the
\usepackage{epsfig}
\usepackage{amssymb}
\usepackage{amsmath}
\usepackage{color}
\usepackage{amsfonts}

\newtheorem{theorem}{Theorem}

\newtheorem{definition}{Definition}
\newtheorem{lemma}{Lemma}

\newtheorem{remark}{Remark}

\newtheorem{assumption}{Assumption}

\newtheorem{problem}{Problem}
\newtheorem{proof of Theorem 1}{Proof of Theorem 1}
\newcommand*{\QEDA}{\hfill\ensuremath{\square}}

\begin{document}

\begin{frontmatter}

\title{Online distributed algorithms for mixed equilibrium problems in dynamic environments} % Title, preferably not more

\thanks[footnoteinfo]{
\emph{Corresponding author: Kaihong Lu.}}

\author[a]{Hang Xu}\ead{xhzs9497@163.com},~   % Add the
\author[a]{Kaihong Lu}\ead{khong\_lu@163.com},~
\author[b]{Yu-Long Wang}\ead{yulongwang@shu.edu.cn},~
\author[c]{Qixin Zhu}\ead{bob21cn@163.com}
\address[a]{College of Electrical Engineering and Automation, Shandong University of Science and Technology,
Qingdao 266590, China}
\address[b]{School of Mechatronic Engineering and Automation, Shanghai University, Shanghai 200444, China}
\address[c]{School of Mechanical Engineering, Suzhou University of Science and Technology, Suzhou 215009, China}

%\address[b]{School of Mechanical Engineering, Suzhou University of Science and Technology, Suzhou 215009, China}  % Please supply
            % full addresses
     % here.

                             % keyword list or with the
                                          % help of the Automatica
                                          % keyword wizard

\begin{abstract}                          % Abstract of not more than 200 words.
In this paper, the mixed equilibrium problem with coupled inequality constraints in dynamic environments is solved by employing a multi-agent system, where each agent only has access to its own bifunction, its own constraint function, and can only communicate with its  immediate neighbors via a time-varying digraph. At each time, the goal of agents is to cooperatively find a point in the constraint set such that the sum of local bifunctions with a free variable is non-negative. Different from existing works, here the bifunctions and the constraint functions are time-varying and only available to agents after decisions are made. To tackle this problem, first, an online
distributed algorithm involving accurate  gradient information is proposed based on mirror descent algorithms and primal-dual strategies. Of particular interest is that dynamic regrets, whose offline benchmarks are to find the solution at each time, are employed to measure the performance of the algorithm.
Under mild assumptions on the graph and the bifunctions, we prove that if the deviation in the
solution sequence grows within a certain rate, then both the dynamic regret and the violation of coupled inequality
constraints increase sublinearly. Second, considering the case where each agent only has access to a noisy estimate
on the accurate gradient, we propose an online distributed algorithm involving the stochastic gradients. The result shows that under the same conditions
as in the first case, if the noise distribution satisfies the sub-Gaussian condition, then dynamic regrets, as
well as constraint violations, increase sublinearly with high probability.
Finally, several simulation examples are presented to corroborate the validity of our results.
\end{abstract}

\end{frontmatter}

% OR

%\begin{figure}
%\begin{center}
%\epsfig{file=jcaesar,width=7cm}
%\caption{Gaius Julius Caesar, 100--44 B.C.}
%\label{fig1}
%\end{center}
%\end{figure}

\section{Introduction}
%In recent years, cooperative control of multi-agent systems (MASs) has been extensively studied, which finds various applications in areas like mobile vehicle cooperative formation, satellite clusters attitude alignment, multiple robots flocking, and so on.
%Many basic and important issues have been studied in cooperative control of MASs,
%including consensus, formation control, stabilizability, and topology selection, to name a few.

In Equilibrium problems (EPs), one shoud find a point to guarante the
non-negativity of a bifunction,  which takes two arguments as inputs and produces a single output. More specifically, one of the two argements is freely changeable.
In recent years, EPs have received increasing attention \cite{htx1}-\cite{htx1.3}. This is due to their wide applications in many fields such as electrical networks \cite{htx2}, transportation networks \cite{htx3} and learning systems \cite{htx4}. For example, to analyze ampere-voltage characteristics of circuits formed by the interconnection of electrical devices like generators, diodes
and inductors, currents in $m$ loops are usually computed by finding $x\in\Omega$ such that
\begin{eqnarray}\label{1}
\langle Ax+b,y-x\rangle+F(y)-F(x)\geq 0, ~\forall y\in\Omega
\end{eqnarray}
where $\Omega\in \mathbb{R}^m$, $F(\cdot): \mathbb{R}^m\rightarrow \mathbb{R}$ is the electrical potential of diodes and is usually nonlinear, $A\in \mathbb{R}^{m\times m}$ and $b\in\mathbb{R}^m$ are determined
by the resistance and voltage source in the circuit, respectively. Problem (1) is a typical EP. If the bifunction is formed by the sum of multiple bifunctions, then such an EP becomes a mixed equilibrium problem (MEP). To achieve a solution of EPs or MEPs, various centralized strategies have been proposed \cite{htx1-1}-\cite{htx1-5}.

In practical applications such as large-scale electrical networks \cite{htx5} and mobile ad-hoc networks \cite{htx5-1}, achieving the global information is usually impossible. Thus, distributed strategies are more desired. Recently, some distributed algorithms have been proposed for MEPs. For example, for MEPs with a set constraint, a distributed extragradient algorithm is proposed in \cite{htx6}. For the case with multiple set constraints, a distributed primal-dual algorithm involving a fixed step size is proposed in \cite{htx7}. To reduce the communication cost, an event-triggered distributed strategy is developed in \cite{htx9}. Moreover, consider the case where the cut property of bifunctions is absent,  an auxiliary optimization-based distributed ergodic algorithm is presented in \cite{htx10}.

If the bifunction in an MEP is time-varying and only available to agents after decisions are made, then the MEP is called
an online MEP. Accordingly, an algorithm for solving such an MEP is referred to as an online algorithm. It is noticed that in all the aforementioned works, MEPs are offline, and feasible sets are only constrained by convex sets.  On the one hand, dynamic environments arise in many practical cases. For example, in problem (1), due to the conductivity of some electronic components such as diodes and inductors, is often influenced by temperature and air pressure, the bifunction in the problem is time-varying, and changes can only be seen in hindsight.  On the other hand,  in practical applications such as wireless sensor networks \cite{htx12} and plug-in electric vehicles \cite{htx13}, coupled inequality constraints are often used to describe some complex and limited resource constraints. Motivated by these  observations, we try to study online MEPs with coupled inequality constraints in a distributed manner. Unfortunately, the results and analysis tools for offline MEPs are not applicable to the online cases.
In offline MEPs, the asymptotic convergence of an algorithm to a fixed solution is proved by using supermartingale
convergence theory \cite{htx6}. However, in online cases, the solution is time-varying and the deviation in the solution
sequence may not be bounded, then the supermartingale convergence theorem is not applicable. A popular and common way for addressing online problems is the regret, and the effectiveness of online algorithms is usually measured by the sublinearity of the regret \cite{htx15}-\cite{htx17-2}.

In fact, online MEPs are also related to online distributed optimization. Specifically, when the bifunction is in the form of the difference between two same functions,  MEPs are reduced to distributed optimization \cite{htx14}.  Recently, online distributed optimization has been extensively studied \cite{tx9}-\cite{htx17}. For example,  online distributed optimization problems without constraints are studied in \cite{tx9} and  \cite{tx10}. For the cases with set constraints, an online distributed saddle point algorithm is proposed in \cite{tx11}, an online distributed mirror descent algorithm is developed in \cite{tx12}, and an online distributed dual averaging algorithm is designed in \cite{tx13}. With a linear inequality constraint considered, an online distributed algorithm based on a safety parameter estimation is proposed in \cite{tx14}. Considering the case with nonlinear inequality constraints,  an online distributed projection-based primal-dual algorithm is presented in \cite{tx15}. Moreover, to address coupled inequality constraints, various online distributed algorithms based on primal-dual strategies are developed in \cite{htx15}-\cite{htx17}. Different from distributed optimization where the goal of agents is to minimize a single-variable function, in MEPs, agents needs to seek a point in a constraint set to ensure that the bifunction is nonnegative.  It is worth pointing out that the bifunctions are coupled through a pair of variables, one of which is freely changeable. More significantly,  the two variables may be inseparable, and the bifunctions may be nonlinear with respect to the two variables.  These factors bring challenges in solving MEPs.

In this paper, online MEPs with coupled inequality constraints are studied by employing a multi-agent system
(MAS). Each agent only has access to information associated with its own bifunction, a set constraint and a local constraint function, and can only communicate with its immediate neighbors via a time-varying digraph. Moreover, the information of local bifunctions is not available to agents until decisions are made. The main novelties and contributions of this paper are summarized as follows.

1) Different from existing works on offline MEPs \cite{htx6}-\cite{htx10}, where only convex set constraints are considered,  here the online case with coupled inequality  constraints is studied.  For online MEPs with coupled inequality constraints, it is necessary for agents to cooperatively estimate the common optimal dual variable when making decisions. But it is unnecessary in the cases with local constraints. So the formulated problem is much more difficult than those studied in \cite{htx14}, \cite{tx14}, \cite{tx15}. Moreover, online MEPs also cover distributed optimization problems \cite{tx9}-\cite{htx17-1} and  distributed Nash equilibrium seeking problems \cite{htx17-2}, \cite{htx35}, \cite{htx36}. To the best of our knowledge, this is the first time to study online MEPs. We are committed to studying this problem under the framework of dynamic regrets, whose offline benchmark is to find the solution to the MEP at each time. Furthermore, we use the deviation in the solution sequence to characterize the difficulty in achieving a sublinear bound of dynamic regrets in the worst case.

2) To handle this problem, we propose a new
online distributed algorithm, called the online distributed mirror descent-based primal-dual algorithm. In this algorithm, two
consensus strategies are involved. One is used to estimate a common dual variable for coupled inequality constraints, while another is employed to cooperatively find the solution of the MEP at each time.   In contrast to the offline continuous-time distributed algorithm proposed in \cite{htx14}, we study the online case, which is applicable to MEPs in dynamic environments. Furthermore, the convergence of the offline algorithm in \cite{htx14} is analyzed by Lyapunov theory and La Salle's invariance principle. However, in online setting, the solution is time-varying and the deviation in the solution sequence may not be bounded, then the analytical methods in \cite{htx14} is not applicable. In this paper, the performance of online algorithms is measured by regrets, whose offline benchmarks are to find the solution at each time. Based on the monotonicity-like property of the bifunctions, combining online optimization theory, mirror descent analysis theory and convex ananlysis theory, we propose a new regret analysis technique for online MEPs. We prove that if the graph is uniformly strongly connected and balanced, and if the deviation in the solution sequence is within a certain range, then both the dynamic regret
and the constraint violation grow sublinearly.

3) Considering the case where the accurate gradient information of bifunctions and constraint functions is not available, we propose an online distributed stochastic gradient algorithm. When implementing the algorithm, agents make decisions only relying on the stochastic gradient rather than the accurate gradient information. Different from \cite{htx18}-\cite{htx20} where the sublinear bound of regrets is analyzed in expectation, we study the sublinear bound of regrets in high probability.  Different from \cite{htx17-1} and \cite{htx17-2}, where
the high probability bounds are only determined by the difference between the stochastic gradient and the accurate gradient, here the high probability
bounds of regrets are simultaneously influenced by the consensus error, the variation of states and the difference between the stochastic gradient and the accurate gradient. This brings more difficulties in analyzing high probability bounds. Using sub-Gaussian analysis techniques and stochastic inequality theory, under the same conditions as those in the previous case, we prove that the dynamic regret and the constraint violation increase sublinearly with high probability.

This paper is organized as follows. In Section \ref{se1}, we formulate the problem to be studied.  In Section \ref{se2}, we propose an online distributed algorithm involving accurate gradient information.  In Section \ref{se2}, we present an online distributed algorithm involving stochastic gradient information. In Section \ref{se4}, simulation examples are presented to demonstrate the validity of our methods. In Section \ref{se5}, we conclude the whole paper.

{\emph{\textbf{Notations}}}.  Throughout this paper, we use $|a|$ to represent the absolute value of scalar $a$. $\mathbb{N}$ is used to represent the set of positive integers. For any $T\in\mathbb{N}$, we denote set $\lfloor T\rfloor=\{1, \cdots, T\}$. $\mathbb{R}^{m}$ and $\mathbb{R}_+^m$ denote $m$-dimensional real vector space and $m$-dimensional non-negative real vector space, respectively. For a given vector ${{x}}\in\mathbb{R}^{m}$, $ \Vert{{x}} \Vert$ denotes the standard Euclidean norm of ${x}$, i.e., $ \Vert{{x}} \Vert=\sqrt{{{x}}^T{{x}}}$. Let ${x}=[x^1, \cdots, x^m]^T$, we denote $[{x}]_+=[\max(x^1, 0), \cdots, \max(x^m, 0)]^T$. For any two vectors ${u},{v}\in \mathbb{R}^m$, the operator $\langle{u}, {v}\rangle$ denotes the inner product of ${u}$ and ${v}$.   Given a function $g(\cdot): \mathbb{R}^m\rightarrow\mathbb{R}$, we use $\nabla g({x})$ to represent its gradient at ${x}\in\mathbb{R}^m$. Given a bifunction $f(\cdot,\cdot): \mathbb{R}^m\times\mathbb{R}^m\rightarrow\mathbb{R}$, we use $\nabla_2 f(\cdot,\cdot)$ to represent its gradient with respect to the second argument. Given a set of vectors $x_i\in\mathbb{R}^{m}$, where $i\in\mathcal{V}$ and $\mathcal{V}=\{1,\cdots,n\}$, we denote $col\{x_i\}_{i\in\mathcal{V}}=[x_1^T,\cdots,x_n^T]^T$. ${1}_m$ denotes the $m$-dimensional vector with elements being all ones and $I_m$ represents an $m\times m$ identity matrix. Given a matrix ${A}\in\mathbb{R}^{m\times n}$, $[{A}]_{ij}$ denotes the matrix entry in the $i${th} row and $j${th} column, $[{A}]_{i\cdot}$ represents the $i${th} row of matrix ${A}$. $\mathcal{O}(T^a)$ means that the order of a function with respect to $T$ is $a$.

\section{Problem formulation}\label{se1}
\subsection{Basic graph theory}
Now let us recall the definition of the uniformly strong connectivity associated with a time-varying directed communication graph.
\begin{definition}
Given a time-varying digraph $\mathcal{G}(t)=(\mathcal{V},\mathcal{E}(t), {A}(t))$ for $t=0, 1, \cdots$, where $\mathcal{V}=\{1, \cdots, n\}$ is the set of nodes,  $n$ is the number of agents satisfying $n\geq 2$, $\mathcal{E}(t)\subset\mathcal{V}\times\mathcal{V}$ is the set of edges, and ${A}(t)=(a_{ij}(t))_{n\times n}$ is the weighted matrix of edges such that $1\geq a_{ij}(t)\geq\gamma$ for some $\gamma>0$ if $(j,i)\in \mathcal{E}(t)$ and $a_{ij}(t)=0$ otherwise. $\mathcal{N}_i(t)=\{j\in\mathcal{V}|(j,i)\in\mathcal{E}(t)\}$ denotes the neighbor set of node $i$ and $i\in\mathcal{N}_i(t)$ for any $i\in\mathcal{V}$. $\mathcal{G}(t)$ is strongly connected if there exists a directed path between each pair of nodes. For $\mathcal{G}(t)$, define the $U$-edge set as $\mathcal{E}_U(t)=\cup_{k=tU}^{(t+1)U-1}\mathcal{E}(k)$ for some $U>0$. If $\mathcal{G}(t)$ with $\mathcal{E}_B(t)$ is strongly connected for any $t\geq0$, then $\mathcal{G}(t)$ is called a $U$-strongly connected (uniformly strongly connected) graph.
\end{definition}

\begin{assumption}\label{as1}
Graph $\mathcal{G}(t)$ is $U$-strongly connected and weighted matrix ${A}(t)$ satisfies ${A}(t){1}_n={A}^T(t){1}_n={1}_n$.
\end{assumption}

The connectivity of a time-varying graph in Assumption \ref{as1} is usually adopted in distributed coordination control, which plays an important role in rendering agents to reach a common state \cite{htx16-1}-\cite{htx17-1}, \cite{Nedi2}. For any $t\geq s\geq 0$, we denote:
\begin{eqnarray}\label{eq0}
\left\{\begin{split}
&{\Phi}(t, s)={A}(t-1)\cdots{A}(s+1){A}(s), ~\emph{\emph{if}}~t> s;\\
&{\Phi}(t, s)={I}_n, ~\emph{\emph{if}}~t=s.
\end{split}
\right.
\end{eqnarray}
On the basis of results in \cite{Nedi2}, we have the following lemma.
\begin{lemma}\label{le1}
Under Assumption 1, for any $i, j\in\mathcal{V}$, $t\geq s\geq 0$,
\begin{eqnarray*}\label{eq01}
\left|[{\Phi}(t, s)]_{ij}-\frac{1}{n}\right|\leq \mathcal{C}\lambda^{t-s}
\end{eqnarray*}
where $\mathcal{C}=2\lambda^{-1}\left(1+\gamma^{-(n-1)U}\right)/\left(1-\gamma^{(n-1)U}\right)$ and $\lambda=\left(1-\gamma^{(n-1)U}\right)^{1/((n-1)U)}$.

\end{lemma}

\subsection{Online MEPs with coupled inequality constraints}
Consider an MAS consisting of $n$ agents, labeled by set $\mathcal{V}=\{1,\cdots, n\}$, where agents communicate with their neighbors via graph $\mathcal{G}(t)$. For each agent $i\in\mathcal{V}$, a set of bifunctions is given by $\{f_i^1, \cdots, f_i^T\}$, where $t\in\lfloor T\rfloor$, $T$ is the finite time horizon, and $f_{i}^t(\cdot, \cdot):\Omega\times\Omega\rightarrow\mathbb{R}$ is convex with respect to the second argument in convex set $\Omega\subset \mathbb{R}^m $. Moreover, the feasible set at each time is constrained by a set of coupled inequalities
\begin{eqnarray}\label{eq02}\begin{split}
\mathcal{{X}}^t=\left\{{x}\in \Omega\Big|g^t(x)=\sum_{i=1}^ng_i^t({x})\leq0\right\}
\end{split}\end{eqnarray}
 where $g_{i}^t({x})=[g_{i1}^t({x}), \cdots, g_{ih}^t({x})]^T$ is convex  and $g_{ik}^t(\cdot):\Omega\rightarrow\mathbb{R}$ for any $i\in\mathcal{V}$ and $k\in\{1, \cdots, h\}$. At each time $t\in\lfloor T\rfloor$, agent $i$ selects a state $x_i(t)\in\Omega$. After the state is selected, the information of $f_{i}^t$ and $g_{i}^t$ is revealed to agent $i$. That is, the agents make decisions at every time only allowing to use the information at the previous time. At each time, the goal of agents is to cooperatively solve the following MEP:
\begin{eqnarray}\label{eq2}\begin{split}
&\emph{\emph{Find}}~~{x}^*_t\in \mathcal{{X}}^t~\emph{\emph{such~that}}\\
& ~f^t({x}^*_t, {y})=\sum_{i=1}^nf_i^t({x}^*_t, {y})\geq0,~\forall {y}\in \mathcal{{X}}^t
\end{split}\end{eqnarray}
where $\mathcal{{X}}^t$ is defined in (\ref{eq02}).

\begin{remark}
Let $f_i^t({x}, {y})=\psi_i^t(y)-\psi_i^t(x),~\forall i\in\mathcal{V}$ for some time-varying function $\psi_i^t(\cdot):\mathbb{R}^m\rightarrow \mathbb{R}$.  In this case, variables
$x$ and $y$  are separable. It is obvious that finding ${x}^*_t\in \mathcal{{X}}^t$  such that $\sum_{i=1}^n(\psi_i^t(y)-\psi_i^t({x}^*_t))\geq0,~\forall {y}\in \mathcal{{X}}^t$ is equivalent to solving $\min_{{x}\in \mathcal{{X}}^t}\sum_{i=1}^n\psi_i^t({x}))$. Accordingly, MEP (\ref{eq2}) is reduced to be an online distributed optimization problem studied in  \cite{htx15}-\cite{htx17}. Moreover, if $f_i^t({x}, {y})=\langle\varphi_i^t(x),y-x\rangle$ for some $\varphi_i^t(\cdot):\mathbb{R}^m\rightarrow \mathbb{R}^m$,  then MEP (\ref{eq2}) is reduced to be a dynamic variational inequality problem. Moreover, Nash equilibrium problems in noncooperative games can usually be formulated as variational inequality problems \cite{htx36}. It is worth noting that convex optimization problems and variational inequality problems can be transformed into MEPs,  but not vice versa. Thus, MEPs are mathematically more general than distributed optimization problems and variational inequality problems. In particular, compared with MEPs studied in \cite{htx6}-\cite{htx10}, we study a more general case where the feasible set is constrained by coupled inequalities and the bifunctions are time-varying.
\end{remark}

Some necassary assumptions are made as follows.

\begin{assumption}\label{as2} For any $i\in\mathcal{V}$, $k\in\{1, \cdots, h\}$ and $t\in\lfloor T\rfloor$,
the following conditions are satisfied:\\
(i) $f_{i}^t(\cdot, \cdot)$ is convex with respect to the second argument and $g_i^t(\cdot)$ is convex;\\
(ii) $\mathcal{X}^t$ is non-empty and $\Omega$ is  compact, i.e., $\forall x\in\Omega$, $\|{x}\|\leq \kappa$ for some $\kappa>0$;\\
(iii) $\|\nabla_2 f_{i}^t({x}, {y})\|\leq \kappa_1$, $\|g_{i}^t({x})\|\leq \kappa_2$ and $\|\nabla g_{ik}^t({x})\|\leq \kappa_3$, $\forall {x}, {y}\in\Omega$ for some $\kappa_1, \kappa_2, \kappa_3>0$.\\
\end{assumption}

 In Assumption 2, condition (i) requires the convexity of $\nabla_2 f^t_i(x,\cdot)$ and $\nabla g^t_i(\cdot)$. Condition (ii) guarantees the existence of the solution. While condition (iii) implies that $f^t_i(x,\cdot)$ and $g_{i}^t(\cdot)$ is differentiable on $\Omega$. The assumption on the convexity and the differentiability are commonly used in the study of online distributed optimization \cite{htx15}-\cite{htx17}. Letting $\sum_{i=1}^nf_i^t({x^*_t}, {y})=\sum_{i=1}^n\psi_i^t(y)-\sum_{i=1}^n\psi_i^t(x^*_t),~\forall {y}\in \mathcal{{X}}^t$, the fact $\sum_{i=1}^nf_i^t({x^*_t}, {y})\geq 0$ implies that $x^*_t$ is the minimizer of $\sum_{i=1}^n\psi_i^t$. To find $x^*_t$, we only need to minimize $\sum_{i=1}^n\psi_i^t$. Obviously, this is a distributed optimization problem,  and the boundedness of gradients is commonly used in distributed optimization. In this scenario, $\nabla\psi_i^t(y)=\nabla_2 f^t_i(x^*_t,y)$. Thus, the boundedness of $\nabla_2 f^t_i(\cdot,\cdot)$ is necessary.
Moreover, the following assumption is made for the bifunctions.

\begin{assumption}\label{as3} For any $i\in\mathcal{V}$, $k\in\{1, \cdots, h\}$ and $t\in\lfloor T\rfloor$,
 the bifunctions satisfy the following conditions:\\
(i) $f_i^t({x}, {x})=0$, $\forall {x}\in\Omega$;\\
(ii) $|f_{i}^t({x}, {z})-f_{i}^t({y}, {z})|\leq L\|{x}-{y}\|$, $\forall {x}, {y}, {z}\in\Omega$ for some $L>0$;\\
(iii) The global bifunction $f^t$ satisfies $f^t({x}, {y})+f^t({y}, {x})\leq 0$ for any ${x}, {y}\in \mathcal{X}^t$, and the equation holds only if $x=y$.
\end{assumption}
If we let $f_i^t({x}, {y})=\psi_i^t(y)-\psi_i^t(x)$ or $f_i^t({x}, {y})=\langle\varphi_i^t(x),y-x\rangle$ for some $\psi_i^t(\cdot):\mathbb{R}^m\rightarrow \mathbb{R}$ and $\varphi_i^t(\cdot): \mathbb{R}^m\rightarrow \mathbb{R}^m$, then the MEP is reduced to be either an optimization problem or a
variational inequality problem. Obviously, condition (i) is satisfied. Condition (ii) implies that $f_{i}^t(\cdot, \cdot)$ is Lipschitz continuous with respect to the first argument. It is easy to verify that condition (ii) is satisfied if $\psi_i(\cdot)$ or $\varphi_i(\cdot)$ is Lipschitz continuous.  Furthermore, if $\varphi_i^t(\cdot)$ is strictly monotonous, i.e., $\langle \varphi_i^t(x)-\varphi_i^t(y),x-y\rangle\geq 0, \forall x,y\in\mathbb{R}^m$ for some $\mu>0$, and the equation holds only if $x=y$, then condition (iii) is satisfied. Hence, condition (iii) can be viewed as the strict monotonicity-like condition, which guarantees the uniqueness of the solution to MEPs \cite{htx1-2}, \cite{htx1-3}, \cite{htx6}.  Accordingly, there exists a unique solution $x^*_t$ such that $f^t({x}^*_t, {x}_0)\geq0$ for any ${x}_0\in \mathcal{{X}}^t$. By condition (iii), there holds $-f^t({x}_0, {x}^*_t)\geq0$ for any ${x}_0\in \mathcal{{X}}^t$. Note that if $f^t({x}_0, {x}^*_t)=0$, then $f^t({x}^*_t, {x}_0)=0$ and $f^t({x}^*_t, {x}_0)+f^t({x}_0, {x}^*_t)=0$. It follows from condition (iii) that $f^t({x}_0, {x}^*_t)=0$ if and only if ${x}_0={x}^*_t$. In online cases, the performance of online algorithms should be measured by the regret. Motivated by the above analysis, the following regret, whose offline benchmark is to find $x_0$ such that $-f^t({x_0}, {x}^*_t)=0$ at each time,  is employed 
\begin{eqnarray}\label{eq200-1}\begin{split}
\mathcal{R}_{i, T}=-\sum_{t=0}^{T}f^t({x}_i(t), {x}^*_t), ~i\in\mathcal{V}
\end{split}\end{eqnarray}
where ${x}^*_t$ is the solution to MEP (4).
Furthermore, we define the violation of the coupled inequality constraints as follows
\begin{eqnarray}\label{eq200-2}
\mathcal{R}^g_{i, T}=\left\|\left[\sum_{t=0}^T\sum_{j=1}^ng_j^t({x}_i(t))\right]_+\right\|, ~i\in\mathcal{V}.
\end{eqnarray}
An online algorithm performs well if regret (\ref{eq200-1}) and violation (\ref{eq200-2}) increase sublinearly, i.e., $\lim_{T\rightarrow \infty}\mathcal{R}_{i,T}/T=0$ and $\lim_{T\rightarrow \infty}\mathcal{R}_{i,T}^g/T=0$. Note that $-f^t({x}_0, {x}^*_t)=0$ implies that ${x}_0={x}^*_t$ for any $t\in\lfloor T\rfloor$, accordingly, the offline benchmark is to seek the solution of MEP (\ref{eq02}) at every time, and regret (\ref{eq200-1}) is dynamic. However, using
dynamic regrets causes the problem unsolvable in the worst case, that is, the sublinearity of the dynamic regret may not be guaranteed if the solution changes fast. Motivated by \cite{htx15}, \cite{htx16}, we characterize the difficulty by using the deviation in the solution sequence
\begin{eqnarray}\label{eq200-3}\begin{split}
{\Theta}_T=\sum_{t=0}^{T}\|x^*_{t+1}-x^*_{t}\|.
\end{split}\end{eqnarray}

\begin{problem}
Suppose that agent $i$ communicate with its neighbors via a time-varying graph $\mathcal{G}(t)$, and only has access
to the information associated with $f_i^t$ and $g_i^t$ in hindsight for any $i\in\mathcal{V}$ and $t\in\lfloor T\rfloor$. The goal of this paper is to design online
distributed strategies for agents to ensure that both the regret in (\ref{eq200-1}) and the violation in (\ref{eq200-2}) grow sublinearly if ${\Theta}_T$ in (\ref{eq200-3}) is within a certain range.
\end{problem}

\section{Online distributed strategy involving accurate gradient information}\label{se2}

\subsection{The design of online distributed algorithm}
Before presenting our algorithm, we introduce the Bregman function as follows \cite{htx30}
\begin{eqnarray}\label{eq6-1}
 \mathcal{D}_\phi({x}, {y})=\phi({x})-\phi({y})-\langle\nabla \phi({y}),{x}-{y}\rangle
\end{eqnarray}
where $\phi(\cdot): \mathbb{R}^m\rightarrow\mathbb{R}$ represents a $\mu$-strongly convex function for some $\mu>0$, i.e.,
\begin{eqnarray*}
 \phi({x})-\phi({y})\geq\langle\nabla \phi({y}), {x}-{y}\rangle +\frac{\mu}{2}\|{x}-{y}\|^2.
\end{eqnarray*}
Here we make the following assumption for $\mathcal{D}_\phi(\cdot, \cdot)$.
\begin{assumption}\label{as4}$\mathcal{D}_\phi(\cdot, \cdot)$ is assumed to satisfy the following conditions:\\
(i) $\mathcal{D}_\phi(\cdot, \cdot)$ is convex with respect to the second argument;\\
(ii) There exists some constant $\ell>0$ such that $\mathcal{D}_\phi(x, z)-\mathcal{D}_\phi(y, z)\leq \ell\|x-y\|$, $\forall x, y, z \in\Omega$;\\
(iii) $\mathcal{D}_\phi(x, y)\leq K\|x-y\|^2$ for some $K>0$.
\end{assumption}

One choice of Bregman functions satisfying Assumption 4 is the Mahalanobis distance $\mathcal{D}_\phi(x,y)=(x-y)^TP^{-1}(x-y)^T$ for some symmetric and positive definite matrix $P \in \mathbb{R}^{m\times m}$. Particularly, if $P=I_m$, then the Bregman function becomes the standard Euclidean distance function $\mathcal{D}_\phi(x,y)=\|x-y\|^2$. Moreover, let the $i${th}  element of vector $\alpha$ be $\alpha^i$, consider $x,y\in \{\alpha\in \mathbb{R}^{m}| 1_m^T\alpha=1, \alpha^i\geq \frac{1}{K}\}$, the Kullback-Leibler function
$\mathcal{D}_\phi(x,y)=\sum_{i=1}^m x^i(\ln x^i-\ln y^i)$ generated by $\phi(x)=\sum_{i=1}^m (x^i\ln x^i-x^i)$ is another special case of Bregman function satisfying Assumption 4.
In fact, Assumption 4 is also adopted in \cite{tx12}.
Note that if $\nabla\phi$ is Lipschitz continuous on $\Omega$, then conditions (ii) and (iii) in Assumption 4 can be satisfied. By (8), we have $\|\nabla_1\mathcal{D}_\phi(x,y)\|\leq \|\nabla\phi(x)\|+\|\nabla\phi(y)\|$. The continuity of $\nabla\phi$ on the compact set $\Omega$ implies that $\|\nabla\phi\|\leq \frac{\ell}{2}$ for some $\ell>0$, then $\|\nabla_1\mathcal{D}_\phi(x,y)\|\leq \ell$ and condition (ii) is satisfied. Letting  $\nabla\phi$ be $K$-Lipschitz continuous for some $K>0$,  together with  the convexity of $\phi$, there holds $\mathcal{D}_\phi(x,y) \leq \langle\nabla\phi(x),x-y\rangle-\langle\nabla\phi(y),x-y\rangle\leq K\|x-y\|^2$, then condition (iii) is satisfied. In fact, it is easy to satisfy the Lipschitz continuity of  $\nabla\phi$ for a strongly convex function $\phi$. For example, consider $\phi(x)=\sum_{i=1}^m (x^i\ln x^i-x^i)$, the Hessian matrix follows that $\nabla^2\phi(x)=\emph{\emph{diag}}\big(\frac{1}{x^1},\cdots, \frac{1}{x^m}\big)$. Letting $K=\|\nabla^2\phi(x)\|=\max_{1\leq i\leq m}|\frac{1}{x^i}|$, then $\nabla\phi$ is $K$-Lipschitz continuous.

Now consider the following optimization problem
\begin{eqnarray}\label{G1}\begin{split}
   &\min_{x\in\Omega}\psi(x)\\
   &s.t.~g(x)\leq0
\end{split}\end{eqnarray}
where $\Omega\subseteq\mathbb{R}^m$ is a closed and convex set; $g(\cdot)=[g_1(\cdot), \cdots, g_h(\cdot)]^T$;  $\psi(\cdot), g_j(\cdot):\mathbb{R}^m\rightarrow\mathbb{R}$ are convex for any $j\in\{1, \cdots, h\}$. To address problem (\ref{G1}), motivated by \cite{htx28}, the regulated Lagrange function is constructed as: $L(x, \lambda)=\psi(x)+\lambda^T g(x)-\frac{\gamma_{0}\|\lambda\|^2}{2}$ for some $ \gamma_{0}>0$, where $\lambda$ is the Lagrange multiplier. Note that if  $\gamma_{0}$ is small enough, the regulated Lagrange function $L(x, \lambda)$ approximates to the standard Lagrange function $L_0(x, \lambda)=\psi(x)+\lambda^Tg(x)$. Based on the regulated Lagrange function $L(x, \lambda)$, combining the mirror descent method \cite{htx18} and the primal-dual strategy \cite{htx17} results in the following algorithm
\begin{eqnarray*}\label{G2}\left\{\begin{split}
   x(t+1)=&\arg\min_{x\in\Omega}\{\mathcal{D}_\phi({x}, x(t))+\\
   &\langle \gamma_{1}\nabla_x L(x(t),\lambda(t)), x\rangle \}\\
   \lambda(t+1)=&[\lambda(t)+\gamma_{2}\nabla_\lambda L(x(t), \lambda(t))]_+
\end{split}\right.\end{eqnarray*}
where $\gamma_{1}, \gamma_{2}>0$ are two step-sizes, and $x(t)$, $\lambda(t)$ are the estimates of the minimizer and the optimal Lagrange multiplier, respectively. It implies that
\begin{eqnarray}\label{G3}\left\{\begin{split}
   x(t+1)=&\arg\min_{x\in\Omega}\{\mathcal{D}_\phi({x}, x(t))+\langle \gamma_{1}\nabla \psi(x(t))\\
   &+\gamma_{1}\nabla g(x(t))\lambda(t), x\rangle \}\\
   \lambda(t+1)=&[(1-\gamma_{2}\gamma_{0})\lambda(t)+\gamma_{2}g(x)]_+.
\end{split}\right.\end{eqnarray}
Multiplying by $\gamma_{0}$ at both sides of the second equation,  then letting $y(t)=\gamma_{0}\lambda(t)$, $\zeta=\gamma_1$, $\eta=\frac{\gamma_{1}}{\gamma_{0}}=\gamma_{2}\gamma_{0}$ in (\ref{G3}), we have the following strategy
\begin{eqnarray}\label{G4}\left\{\begin{split}
   x(t+1)=&\arg\min_{x\in\Omega}\{\mathcal{D}_\phi({x}, x(t))+\langle \zeta\nabla \psi(x(t))\\
   &+\eta\nabla g(x(t))y(t), x\rangle \}\\
  y(t+1)=&[(1-\eta)y(t)+\eta g(x(t))]_+.
\end{split}\right.\end{eqnarray}
Motivated by (\ref{G4}), to solve MEP (\ref{eq2}), the following online distributed algorithm is proposed
\begin{eqnarray}\label{eq8}
\left\{ {\begin{array}{*{20}c}\begin{split}
   {x}_i(t+1)=&\arg\min_{{x}\in\Omega}\Big\{\mathcal{D}_\phi({x}, z_i(t))+\Big\langle \zeta_t\nabla_2f_i^t(x_i(t),\\
   & {x}_i(t))+\eta_t\nabla g_i^t(x_i(t)){y}_i(t), {x}\Big\rangle\Big\}\\
   y_i(t+1)=&\Big[(1-\eta_t)\sum_{j\in\mathcal{N}_i(t)}a_{ij}(t)y_j(t)\\
   &+\eta_t g_i^t(x_i(t))\Big]_+\\
  {z}_i(t)=&\sum_{j\in\mathcal{N}_i(t)}a_{ij}(t){x}_{j}(t)\\
  \end{split}\end{array}} \right.
\end{eqnarray}
where $x_i(t)$ is the state of agent $i$ with $x_i(0)\in\Omega$, $\nabla g_i^t(\cdot)=[\nabla g_{i1}^t(\cdot),\cdots,$ $\nabla g_{ih}^t(\cdot)]\in\mathbb{R}^{m\times h}$, ${{y}}_i(t)\in \mathbb{R}^h$ is agent $i$'s dual variable with ${{y}}_i(0)=0$, $z_i(t)$ is the estimate on the consensus state of agents, and $\zeta_t, \eta_t\in (0,1]$ are two non-increasing step sizes such that $\zeta_t\leq \eta_t$. Note that when running algorithm (\ref{eq8}), agents make decisions only using local state information, their own bifunctions and constraint functions in the past time, which implies that algorithm (\ref{eq8}) is online and distributed.

\begin{remark}
In algorithm (\ref{eq8}), the design of $y_i(t)$ and $z_i(t)$ is motivated by the consensus algorithm \cite{Nedi2}. Particularly, the boundedness of $y_i(t)$ is guaranteed by regulating the dual strategy.  Different from existing works on distributed optimization, we are committed to investigating the case where $f_i^t(x, y)$ is nonlinear with respect to $x$ and $y$, and these two variables are inseparable. In such a case, $x_i(t)$ cannot be updated by a simple projection algorithm. In this paper, we use the mirror descent algorithm involving the Bregman function to address the nonlinearity of $f_i^t$ and the inseparability of $x$ and $y$.
\end{remark}

\subsection{The convergence analysis}
Before presenting the bounds of regret (\ref{eq200-1}) and violation (\ref{eq200-2}), some necessary lemmas need to be provided. First, a lemma associated with $\mathcal{D}_\phi$ is presented as follows.
\begin{lemma}\label{le2}\cite{htx30}
Consider the following strongly convex optimization problem
\begin{eqnarray*}
\min_{{x}\in\Omega}\left\{\mathcal{D}_\phi({x}, {y})+\langle {s}, {x}\rangle\right\}
\end{eqnarray*}
where $y, s\in \mathbb{R}^m$,  and $\mathcal{D}_\phi$ is defined in (\ref{eq6-1}). $\hat{{x}}$ is the solution to the optimization problem if and only if
\begin{eqnarray*}
 \langle {s},\hat{{x}}-{w}\rangle\leq \mathcal{D}_\phi({w}, {y})-\mathcal{D}_\phi({w}, \hat{{x}})-\mathcal{D}_\phi(\hat{{x}}, {y})
 \end{eqnarray*}
 for any ${w}\in\Omega$.
\end{lemma}

In the next lemma, we show that for any $i\in\mathcal{V}$ and $t\in\lfloor T\rfloor$, $y_i(t)$ is bounded.
\begin{lemma}\label{le4}
Under Assumptions \ref{as1} and \ref{as2}, $\|{y}_i(t)\|\leq{\sqrt{n}\kappa_2}$ for any $i\in\mathcal{V}$ and $t\in\lfloor T\rfloor$, where $\kappa_2$ is defined in Assumption \ref{as2}.
 \end{lemma}

\textbf{Proof.} See Appendix A. \QEDA

To analyze the sublinearity of the dynamic regret, we are now ready to present the bound of $-\zeta_tf^t(x_i(t), {x}^*_t)$.
\begin{lemma}\label{le5}
Under Assumptions \ref{as1}-\ref{as4}, for any $i\in\mathcal{V}$ and $t\in\lfloor T\rfloor$, there holds
  \begin{eqnarray}\label{eq24}
\begin{split}
&-\zeta_t f^t(x_i(t), x^*_t)\\
&\leq-\eta_t\sum_{i=1}^n(y_i(t))^Tg_i(x_i(t))+\kappa_2\eta_t\sum_{i=1}^n\|y_i(t)-\bar{y}(t)\|\\
&~~~+\rho\eta_t\sum_{i=1}^n\|x_i(t)-x_i(t+1)\|\\
&~~~+\sum_{i=1}^n\Big(\mathcal{D}_\phi(x^*_t, x_i(t))-\mathcal{D}_\phi(x^*_{t+1}, x_i(t+1))\Big)\\
&~~~+n\ell\|x^*_{t+1}-x^*_{t}\|+2nL\eta_t\|x_i(t)-\bar{x}(t)\|.
  \end{split}
\end{eqnarray}
where $\rho=\sqrt{n}h\kappa_2\kappa_3+\kappa_1$.
 \end{lemma}

\textbf{Proof.} See Appendix B. \QEDA

Based on the bound in (\ref{eq24}) and the definition of dynamic regret (\ref{eq200-1}), we know
that the bound of the dynamic regret is influenced by terms $\|{x}_i(t)-\bar{x}(t)$, $\|{y}_i(t)-\bar{y}(t)\|$ and $\|{x}_i(t+1)-x_i(t)\|$.
In what follows, we present the convergence bounds of these terms.
\begin{lemma}\label{le6}
Under Assumptions \ref{as1} and \ref{as2}, for any $i\in\mathcal{V}$ and $t\in\lfloor T\rfloor$; \\
(i) $\|{x}_i(t)-\bar{x}(t)\|\leq \rho_1\lambda^{t}+\rho_2\sum_{s=0}^{t}\lambda^{t-s}\eta_{s}$;\\
(ii) $\|{x}_i(t)-\bar{x}(t)\|^2\leq \rho_3\lambda^{t}+\rho_4\sum_{s=0}^{t}\lambda^{t-s}\eta_{s}^2$;\\
(iii) $\|{x}_i(t+1)-x_i(t)\|\leq 2\rho_1\lambda^{t}+2\rho_2\sum_{s=0}^{t}\lambda^{t-s}\eta_{s}+\rho_5\eta_t$;\\
(iv) $\|{y}_i(t)-\bar{y}(t)\|\leq \rho_6\sum_{s=0}^{t}\lambda^{t-s}\eta_{s}$;\\
 (v) $\|y_{i}(t)-\bar{y}(t)\|^2\leq\rho_7\sum_{s=0}^{t}\lambda^{t-s}\eta_{s}^2$\\
where $\bar{x}(t)=\frac{1}{n}\sum_{i=1}^nx_i(t)$, $\bar{y}(t)=\frac{1}{n}\sum_{i=1}^ny_i(t)$, $\rho_1=\sqrt{m}n\kappa\mathcal{C}$,
$\rho_2=\frac{\sqrt{m}{n}^{\frac{3}{2}}h\kappa_2\kappa_3\mathcal{C}+\sqrt{m}n\kappa_1\mathcal{C}}{\mu\lambda}$,
$\rho_3=mn^2\kappa^2\mathcal{C}^2+\frac{2mn^{\frac{5}{2}}h\kappa\kappa_2\kappa_3\mathcal{C}^2+2mn^2\kappa\kappa_1\mathcal{C}^2}{\mu\lambda(1-\lambda)}$,
$\rho_4=\frac{m({n}^{\frac{3}{2}}h\kappa_2\kappa_3\mathcal{C}+n\kappa_1\mathcal{C})^2}{\mu^2\lambda^2(1-\lambda)}$, $\rho_5=\frac{\sqrt{n}h\kappa_2\kappa_3+\kappa_1}{\mu}$
 $\rho_6=\frac{\sqrt{h}n^{\frac{3}{2}}\mathcal{C}\kappa_2+\sqrt{h}n\mathcal{C}\kappa_2}{\lambda}$, and $\rho_7=\frac{h(n^{\frac{3}{2}}\mathcal{C}\kappa_2+n\mathcal{C}\kappa_2)^2}{\lambda^2}$.
\end{lemma}

\textbf{Proof.}  See Appendix C. \QEDA

Note that the bound in Lemma \ref{le5} is also determined by term $-\eta_t(g_i^t(x_i(t)))^Ty_i(t)$. The next lemma shows the bound of the term.
\begin{lemma}\label{le7}
Under Assumptions \ref{as1} and \ref{as2}, there holds
\begin{eqnarray}\label{8-1}\begin{split}
& -\eta_t\sum_{i=1}^n\Big(g_i^t(x_i(t))\Big)^Ty_i(t)\\
&\leq \frac{1}{2}\sum_{i=1}^n\Big(\|y_i(t)\|^2-\|y_i(t+1)\|^2\Big)\\
&~~~~+\eta_t^2(n\kappa_2^2+n^2\kappa_2^2)+\frac{9}{2}\sum_{i=1}^n\|y_i(t)-\bar{y}(t)\|^2\\
&~~~~+2\eta_t(\kappa_2+\sqrt{n}\kappa_2)\sum_{i=1}^n\|y_i(t)-\bar{y}(t)\|.
\end{split}\end{eqnarray}
\end{lemma}
\textbf{Proof.} See Appendix D. \QEDA

With the help of lemmas above, now we present our main result. See the following theorem for details.
\begin{theorem}\label{TH1}
Under Assumptions 1-4, if $\zeta_{t}=(t+1)^{-a}$ and $\eta_{t}=(t+1)^{-b}$ with $a,b \in (0,1)$ and $b<a< 2b$, then by algorithm (\ref{eq8}), for any ${i}\in\mathcal{V}$,
\begin{eqnarray}\label{7-1}\begin{split}
&\mathcal{R}_{i,T}\leq \mathcal{O}\Big({T}^{1+a-2b}+T^{a}\Theta_T\Big)
\end{split}\end{eqnarray}
and
\begin{eqnarray}\label{7-2}\begin{split}
&\mathcal{R}^g_{i,T}\leq \mathcal{O}\Big(\sqrt{{T}^{2+b-a}+T^{1+b}\Theta_T}\Big)
\end{split}\end{eqnarray}
where $R_{i,T}$ and $R_{i, T}^{g}$ are defined in (\ref{eq200-1}) and (\ref{eq200-2}), respectively.
\end{theorem}
\textbf{Proof.} See Appendix E. \QEDA

\begin{remark}
 By Theorem 1, we know that the bounds of the dynamic regret and the violation are influenced by $\Theta_T$. If $\Theta_T< \mathcal{O}(T^{1-a})$, then $\lim_{T\rightarrow\infty}\mathcal{R}_{i,T}/T=0$ and $\lim_{T\rightarrow\infty}\mathcal{R}^g_{i,T}/T=0$, which implies online distributed algorithm (\ref{eq8}) solves MEP (\ref{eq2}) well. If the solution sequence fluctuates drastically, $\Theta_T$ might become linear with respect to $T^{1-a}$, then MEP (\ref{eq2}) becomes insolvable. This is natural since even in distributed optimization, online problems are unsolvable in the worst cases  \cite{htx15}-\cite{htx17}. Different from \cite{htx15}-\cite{htx17}, we study online MEPs with coupled inequality constraints.  Our achieved results are applicable to dealing with more general problems, including distributed optimization problems and variational inequality problems, especially in the scenarios where the functions are determined by two variables.
\end{remark}

\section{Online distributed strategy involving stochastic gradient information}\label{se3}
\subsection{The design of online distributed strategy}
Consider the case where each agent only has access to noisy estimates on  gradients of $f_i^t(\cdot,\cdot)$ and $g_i^t(\cdot)$, denoted by  noisy  gradients
$\nabla_2 F_i^t(\cdot,\cdot,\theta_i^t)$ and $\nabla G_i^t(\cdot,\theta_i^t)$ respectively, where $\theta_i^t$ represents a random variable.  To solve MEP (\ref{eq2}) in  this setting, the online distributed strategy involving stochastic gradient information is proposed as follows
\begin{eqnarray}\label{b1}
\left\{ {\begin{array}{*{20}c}\begin{split}
   {x}_i(t+1)=&\arg\min_{{x}\in\Omega}\Big\{\mathcal{D}_\phi({x}, z_i(t))\\
   &+\Big\langle\zeta b_i^t+\eta c_i^t{y}_i(t), {x}\Big\rangle\Big\}\\
   y_i(t+1)=&\Big[(1-\eta)\sum_{j\in\mathcal{N}_i(t)}a_{ij}(t)y_j(t)\\
   &+\eta g_i^t(x_i(t))\Big]_+\\
  {z}_i(t)=&\sum_{j\in\mathcal{N}_i(t)}a_{ij}(t){x}_{j}(t)\\
  \end{split}\end{array}} \right.
\end{eqnarray}
where $b^t_i=\nabla_2 F_i^t(x_i(t),x_i(t),\theta_i^t)$, $c^t_i=\nabla G_i^t(x_i(t),\theta_i^t)$ and $\zeta,\eta\in (0, 1]$ are two fixed step sizes such that $\zeta\leq \eta$.
It is worth noting that when running algorithm (\ref{b1}),  each agent makes decisions only relying on the noisy gradients of their own bifunctions, as well as the noisy gradients of local constraint functions. Thus, algorithm (\ref{b1}) is applicable to scenarios where the objective functions and constraint functions are stochastic, or accurate gradients are not available \cite{htx22}, \cite{htx23}.

The definition of the high probability bound is presented.
\begin{definition} (High probability bound) For regret $\mathcal{R}_{i,T}$
defined in (\ref{eq200-1}), we call that $\mathcal{R}_{i,T}$  has a high probability bound if
for any failure probability $\nu\in(0,1)$, $\mathcal{R}_{i,T}\leq \mathcal{O}\big(h(T)f(\ln{\frac{1}{\nu}})\big)$ with probability at
least $1-\nu$, where $h(\cdot):\mathbb{R}\rightarrow\mathbb{R}$ and $f(\cdot):\mathbb{R}\rightarrow\mathbb{R}$. Moreover, if
$h(T)$ is sublinear with $T$, we call that $\mathcal{R}_i^d(T)$ increases sublinearly with high probability.
\end{definition}

Define the $\sigma$-field generated by the entire history of the random variables as
\begin{eqnarray*}\begin{split}
\mathcal{F}_t=\{col\{x_i(s)\}_{i\in\mathcal{V}},col\{\theta_i^s\}_{i\in\mathcal{V}}; s=0,\cdots,t-1\}
\end{split}\end{eqnarray*}
with $\mathcal{F}_0=col\{x_i(0)\}_{i\in\mathcal{V}}$.
The noisy gradients are assumed to satisfy the following conditions.
\begin{assumption} \label{as5}
For any $i\in\mathcal{V}$ and $t\in\lfloor T\rfloor$,\\
(i) $\mathbb{E}[\nabla_2 F_i^t(x_i(t),x_i(t),\theta_i^t)|\mathcal{F}_t]=\nabla_2f_i^t(x_i(t),x_i(t))$;\\
(ii) $\mathbb{E}\Big[\exp\Big(\frac{\|\nabla_2 F_i^t(x_i(t),x_i(t),\theta_i^t)-\nabla_2f_i^t(x_i(t),x_i(t))\|^2}{\sigma_1^2}\Big)\Big|\mathcal{F}_t\Big]\leq \exp(1)$ for some $\sigma_1>0$;\\
(iii) $\mathbb{E}[\nabla G_i^t(x_i(t),\theta_i^t)|\mathcal{F}_t]=\nabla g_i^t(x_i(t))$;\\
(iv) $\mathbb{E}\Big[\exp\Big(\frac{\|\nabla  G_i^t(x_i(t), \theta_i^t)-\nabla g_i^t(x_i(t))\|^2}{\sigma_2^2}\Big)\Big|\mathcal{F}_t\Big]\leq \exp(1)$ for some $\sigma_2>0$.\\
\end{assumption}
Conditions (i) and (iii) in Assumption \ref{as5} ensure that noisy gradients are the unbiased estimates on accurate gradients, while conditions (ii) and (iv) in Assumption \ref{as5} describe a sub-Gaussian noise condition, i.e., the tails of the noise distribution are dominated by tails of a Gaussian distribution. Assumption \ref{as5} is commonly used to guarantee the high probability convergence bound in learning systems \cite{htx31}-\cite{htx33}.

\subsection{The convergence analysis}
The high probability bounds of stochastic terms determined by noisy gradients and accurate gradients are presented in the following lemma.
\begin{lemma}\label{le9-1}
Under Assumptions \ref{as2} and \ref{as5}, for any $\nu\in(0,1)$, with probability at least $1-\nu$,
\begin{eqnarray}\label{13}\begin{split}
&\sum_{t=0}^T\sum_{i=1}^n\eta \big\langle b_i^t-\nabla_2f_i^t(x_i(t),x_i(t)),x^*_t-x_i(t)\big\rangle\\
&\leq 2n^\frac{3}{2}\kappa\sigma_1(T+1)\eta^2 +2\sqrt n\kappa\sigma_1\ln\frac{1}{\nu}
\end{split}\end{eqnarray}
and with probability at least $1-\nu$,
\begin{eqnarray}\label{13*}\begin{split}
&\sum_{t=0}^T\sum_{i=1}^n\eta \langle c_i^t{y}_i(t)-\nabla g_i^t(x_i(t)){y}_i(t), {x}^*_t-{x}_i(t)\rangle\\
&\leq 2n^\frac{3}{2}\kappa\sigma_2(T+1)\eta^2 +2\sqrt n\kappa\sigma_2\ln\frac{1}{\nu}.
\end{split}\end{eqnarray}
\end{lemma}
\textbf{Proof.}  See Appendix F. \QEDA

By algorithm (\ref{b1}), we know that agents use noisy gradients to update their states $x_i(t)$. Accordingly, the bounds of terms $\|{x}_i(t)-\bar{x}(t)$ and $\|{x}_i(t+1)-x_i(t)\|$ will be influenced by noisy gradients, and the corresponding results established in Lemma \ref{le6} no longer hold in such a case. In what follows, we present the high probability bound of consensus error term $\|{x}_i(t)-\bar{x}(t)$, as well as the high probability bound of term $\|{x}_i(t+1)-x_i(t)\|$.
\begin{lemma}\label{le10}
Under Assumptions \ref{as1}, \ref{as2} and \ref{as5}, for any $i\in\mathcal{V}$ and $\nu\in(0,1)$, with probability at least $1-\nu$,
\begin{eqnarray}\label{15}\begin{split}
&\sum_{t=0}^T \left\|x_i(t)-\bar{x}(t)\right\|^2\leq \mathcal{O}\Big(T\eta^2+\ln{\frac{1}{\nu}}\Big)
\end{split}\end{eqnarray}
and with probability at least $1-\nu$,
\begin{eqnarray}\label{15*}
\begin{split}
&\sum_{t=0}^T \left\|{x}_i(t)-{x}_i(t+1)\right\|^2\leq \mathcal{O}\Big(T\eta^2+\ln{\frac{1}{\nu}}\Big).
  \end{split}
\end{eqnarray}
\end{lemma}
\textbf{Proof.} See Appendix G. \QEDA

Based on lemmas above, now we present our main result. See the following theorem for details.
\begin{theorem}\label{TH2}
Under Assumptions \ref{as1}-\ref{as5}, if $\zeta =(T+1)^{-a}$ and $\eta =(T+1)^{-b}$ with $a,b \in (0,1)$ and $b<a< 2b$, then by algorithm (\ref{b1}), for any ${i}\in\mathcal{V}$ and  $\nu\in(0,1)$, with probability at least $1-\nu$,
\begin{eqnarray}\label{7-5}\begin{split}
&\mathcal{R}_{i,T}\leq \mathcal{O}\left({T}^{1+a-2b}+T^{a}\Theta_T+T^{a}\ln \frac{1}{\nu}\right)
\end{split}\end{eqnarray}
and
\begin{eqnarray}\label{7-6}\begin{split}
&\mathcal{R}^{g}_{i, T}\leq \mathcal{O}\left(\sqrt{{T}^{2+b-a}+T^{1+b}\Theta_T+T^{1+b}\ln\frac{1}{\nu}}\right)
\end{split}\end{eqnarray}
where $R_{i,T}$ and $R_{i, T}^{g}$ are defined in (\ref{eq200-1}) and (\ref{eq200-2}), respectively.
\end{theorem}
\textbf{Proof.} See Appendix H. \QEDA

\begin{remark}
By Theorem \ref{TH2}, one can find that the high probability bounds of the
dynamic regret and the violation are influenced by $\Theta_T$. If $\Theta_T$ is sublinear with respect to ${T}^{1-a}$, i.e., $\Theta_T<\mathcal{O}({T}^{1-a})$,  then $\mathcal{R}_{i,T}$ and $\mathcal{R}^{g_0}_{i, T}$ have a sublinear bound with high probability, which implies that online stochastic algorithm (\ref{eq8}) performs well.
Moreover,  the bounds of the regret and the violation are also determined by high probability terms $T^{a} \ln \frac{1}{\nu}$ and $T^{1+b} \sqrt{\ln \frac{1}{\nu}}$ respectively. Note that the value of $\ln \frac{1}{\nu}$ increases slowly as the value of failure probability $\nu$ decreases. For example, due to the facts that $\ln 10^2=4.61$, $\ln 10^3=6.91$, $\ln 10^4=9.21$, the term $T^{a} \ln \frac{1}{\nu}$ sublinearly increases as $4.61 T^{a}$, $6.91 T^{a}$, $9.21 T^{a}$, with probabilities at least $99.99\%$, $99.999\%$, $99.9999\%$, respectively. Hence, based on Theorem \ref{TH2}, the sublinearity of the regret and the violation with a probability close to one can be ensured by running algorithm (\ref{b1}) in a single round.
\end{remark}

\begin{remark}
By (\ref{eq26}) and (\ref{b4}), we know that the bounds of the regret and the violation in Theorems 1 and 2 are
also influenced by the network parameter $\lambda$. Since each agent is its own neighbor, the lower bound $\gamma$ of weights is not larger than $\frac{1}{2}$. By the definition of
$\lambda$ in Lemma \ref{le1}, we know that $\frac{1}{2}\leq\lambda <1$ for any $n\geq 2$. It is not difficult to verify that when $\frac{1}{2}\leq\lambda <1$ ,
the value of $\frac{1}{1-\lambda}$ increases as the connected period $U$ and the number $n$ increase. Then, it follows that the bounds of the regret and the violation increase if $U$ and $n$ increase. Consequently, a smaller connectivity period and a smaller number of agents imply a lower bound. Since we only focus on the terms determined by $T$, the network properties do not change the sublinearity of the regret and the constraint violation. Particularly, compared with works \cite{htx17-1}, \cite{htx17-2}  where the communication graph is assumed to be undirected and fixed, the connectivity conditions related to the graph used here are weaker.
\end{remark}

\begin{remark}
For online algorithms, a time-varying step size is usually used to achieve a sublinear bound of the regret. More specifically, if $\eta_t$ is replaced by a positive constant $\eta$, then the term $\sum_{t=0}^T\eta_t^2$ in (\ref{eq26}) becomes $\sum_{t=0}^T\eta^2= (T+1)\eta^2=\mathcal{O}(T)$.
It immediately leads to a linear bound of regrets. If $\eta_t$ is selected to be $\eta_{t}=(t+1)^{-b}$ for some $b \in (0,1)$, then $\sum_{t=0}^T\eta_t^2\leq\mathcal{O}(T^{1-2b})$ and a sublinear bound of regrets can be achieved.  The time-varying step size is commonly adopted in the study of online optimization problems \cite{tx15}-\cite{htx17}. In fact, it is not difficult to use a fixed step size to ensure the sublinearity of regrets. Note that in online problems, the time horizon $T$ is usually considered to be finite. Thus, the sublinearity of regrets can be achieved if the fixed step size is depended on $T$. For example, if the fixed step size $\eta$ is selected to be $\eta =(T+1)^{-b}$ for some $b \in (0,1)$, then $\sum_{t=0}^T\eta ^2\leq\mathcal{O}(T^{1-2b})$, which has the same bound as that in the case with $\eta_{t}=(t+1)^{-b}$. Different from algorithm (\ref{eq8}), the fixed step sizes are adopted in algorithm   (\ref{b1})  to handle the setting where stochastic gradients and coupled inequality constraints simultaneously exist. Note that if $\zeta=\zeta_t$ and $\eta=\eta_t$, then by (\ref{b4-1}), we know that $-\sum_{t=0}^T \zeta_t f^t(x_i(t), x^*_t)\leq\mathcal{O}\big(T^{1-2a}+\Theta_T+\ln\frac{1}{\nu}\big)$. Due to the fact that the non-negativity of $-f^t(x_i(t), x^*_t)$ is unknown, the bound of $-\sum_{t=0}^T f^t(x_i(t), x^*_t)$ can not be analyzed. If we do not consider the stochastic gradients, then the analysis methods follow the proof of Theorem \ref{TH1}. In addition, if the coupled inequality constraints are removed, then  $-f^t(x_i(t), x^*_t)\geq 0$ and the sublinear bound of regrets can be achieved. However, when stochastic gradients and coupled inequality constraints simultaneously exist, the time-varying step sizes fail to guarantee the sublinearity of regrets.  Both stochastic gradients and coupled inequality constraints decrease the convergence rate of the online algorithm and enlarge the bound of the regret. Here we use fixed step sizes to guarantee the effectiveness of the online algorithm. Recently, fixed step sizes are also adopted in \cite{htx33-1}-\cite{htx33-3} for improving the bounds of regrets.
\end{remark}

%\begin{figure}
%\centering
%\includegraphics[width=0.4\textwidth]{lu1.eps}
%\caption{The time-varying digraph sequence.} \label{fig1}
%\end{figure}

\section{Simulation examples}\label{se4}
In this section, we present two numerical examples to illustrate the validity of our theoretical results.

\begin{figure}
\centering
\includegraphics[width=0.4\textwidth]{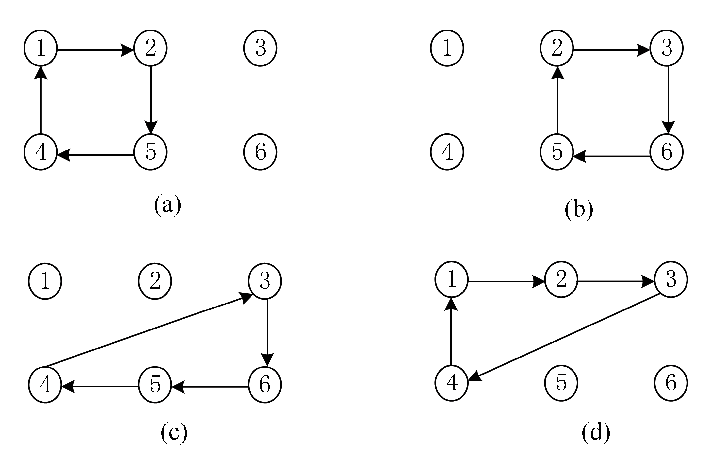}
\caption{The time-varying digraph.} \label{fig1}
\end{figure}

\begin{figure}
\centering
\includegraphics[width=0.4\textwidth]{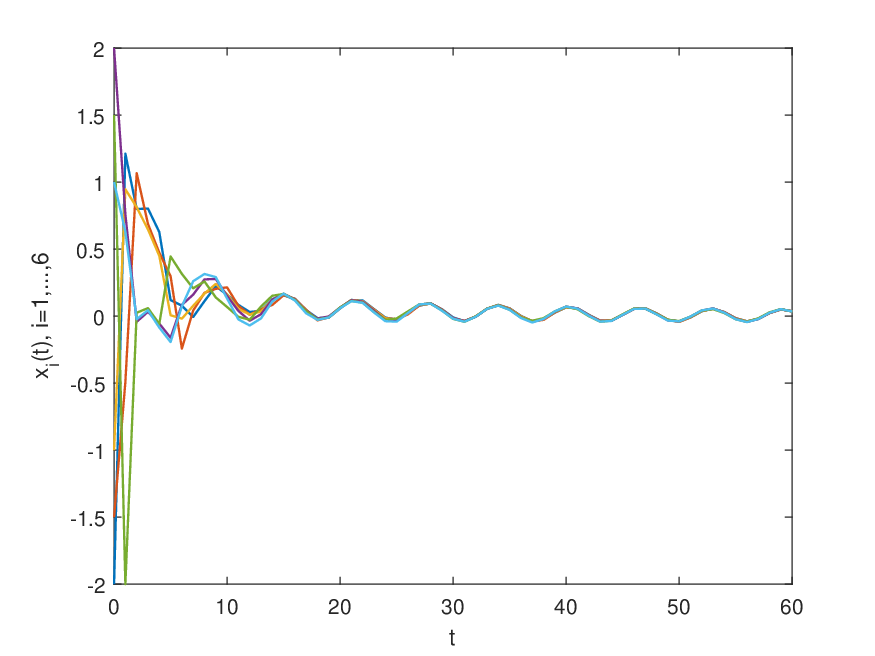}
\caption{The evolutions of $x_i(t)$ under algorithm (\ref{eq8}), $i=1,\cdots,6$.} \label{fig2}
\end{figure}

\begin{figure}
\centering
\includegraphics[width=0.4\textwidth]{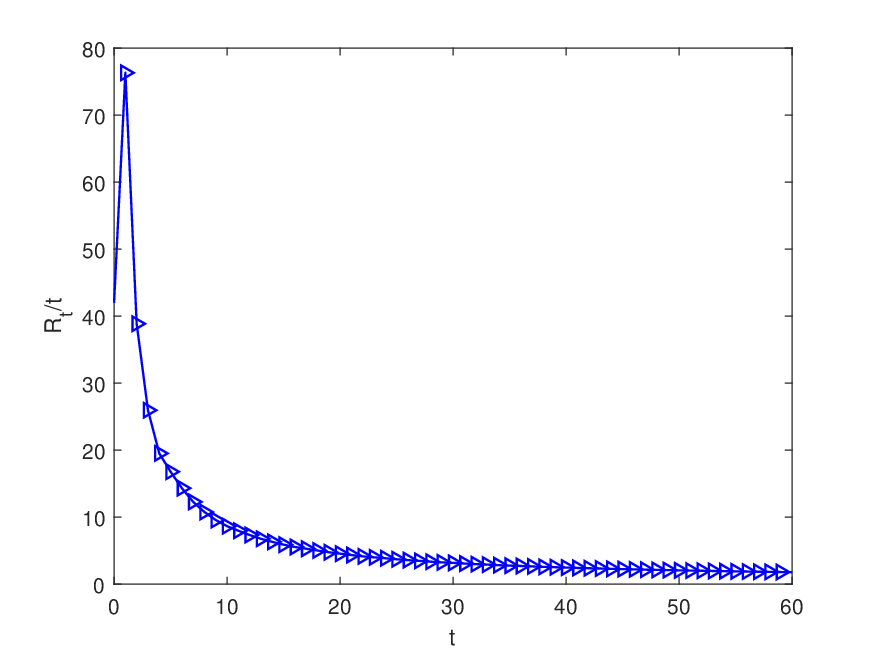}
\caption{The evolution of $\mathcal{R}_{t}/t$ under algorithm (\ref{eq8}).} \label{fig3}
\end{figure}

\begin{figure}
\centering
\includegraphics[width=0.4\textwidth]{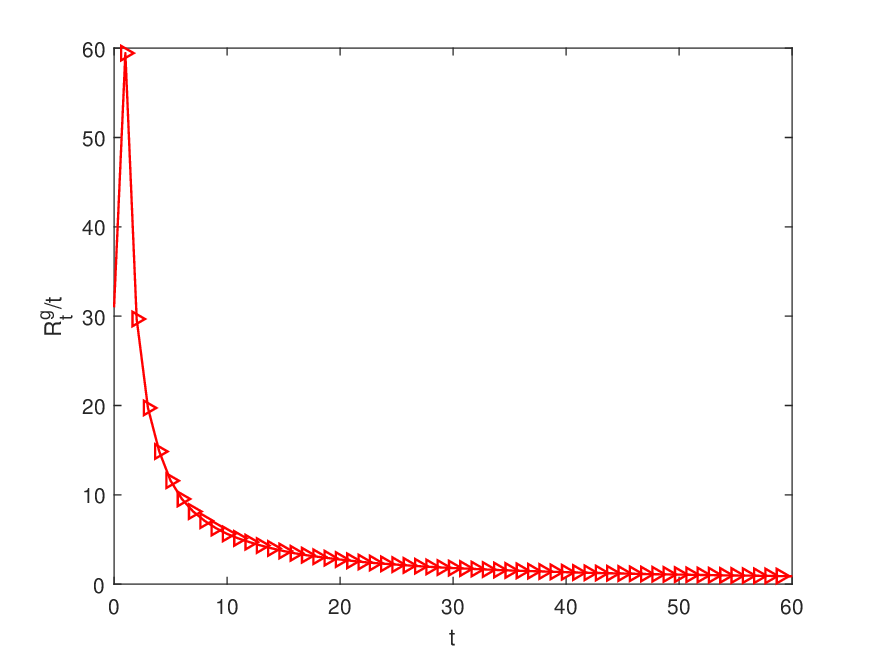}
\caption{The evolution of $\mathcal{R}^g_{t}/t$ under algorithm (\ref{eq8}).} \label{fig4}
\end{figure}

\textbf{Example 1}~
Consider an MAS with six agents, labeled
by set $\mathcal{V}=\{1,\cdots,6\}$. Each agent communicates with its neighbors via a time-varying graph, as shown in Fig. \ref{fig1}, where (a), (b),
(c) and (d) are four graphs in a period with $U=4$.
At each time $t\in \lfloor T\rfloor$, the objective of agents is to cooperatively find a $x^*_t\in\mathcal{{X}}^t$ such that:
\begin{eqnarray*}
\left\{ {\begin{array}{*{20}c}\begin{split}
&\sum_{i=1}^6f_i^t({x}^*_t, {y})\geq0,~\forall{y}\in \mathcal{{X}}^t\\
&f_i^t(x,y)=\frac{i}{2}(y^2-x^2)-3(y-x)\sin t\\
&\mathcal{{X}}^t=\left\{{x}\in \Omega\Big|g^t(x)=\sum_{i=1}^5g_i^t({x})\leq0\right\}
  \end{split}\end{array}} \right.
\end{eqnarray*}
where $\Omega=\{x|-2\leq x\leq2\}$ and $g_i^t(x)=(\sin t+1) x^2-\frac{i}{6}x$.  It is easy to verify that the solution of the MEP is $x^*_t=\frac{6\sin t}{7}$. Now we apply algorithm (\ref{eq8}) to solving the MEP. Let $\zeta_t=1/\sqrt{20t+8}$, $\eta_t=1/(20t+8)^{\frac{1}{3}}$, $\mathcal{D}_\phi({x}, z_i(t))=\|{x}-z_i(t)\|^2$, and the initial states be $x_1(0)=-2, x_2(0)=-1.5, x_3(0)=-1, x_4(0)=2, x_5(0)=1.5$ and $x_6(0)=1$. Define $\mathcal{R}_t=\max_{1\leq i\leq6}\mathcal{R}_{i,t}$ and $\mathcal{R}^g_{t}=\max_{1\leq i\leq6}\mathcal{R}^g_{i,t}$. Figs. \ref{fig2}, \ref{fig3} and \ref{fig4} show the evolutions of $x_i(t)$, $\frac{\mathcal{R}_{t}}{t}$ and $\frac{\mathcal{R}^g_{t}}{t}$, respectively. From Fig. \ref{fig2}, we can see that the states of all agents asymptotically approach to $x^*_t$. Furthermore, observing Fig. \ref{fig3} and Fig. \ref{fig4}, one can find that both $\frac{\mathcal{R}_{t}}{t}$ and $\frac{\mathcal{R}^g_{t}}{t}$ asymptotically decay to zero. These observations illustrate the validity of Theorem 1.


\begin{figure}
\centering
\includegraphics[width=0.35\textwidth]{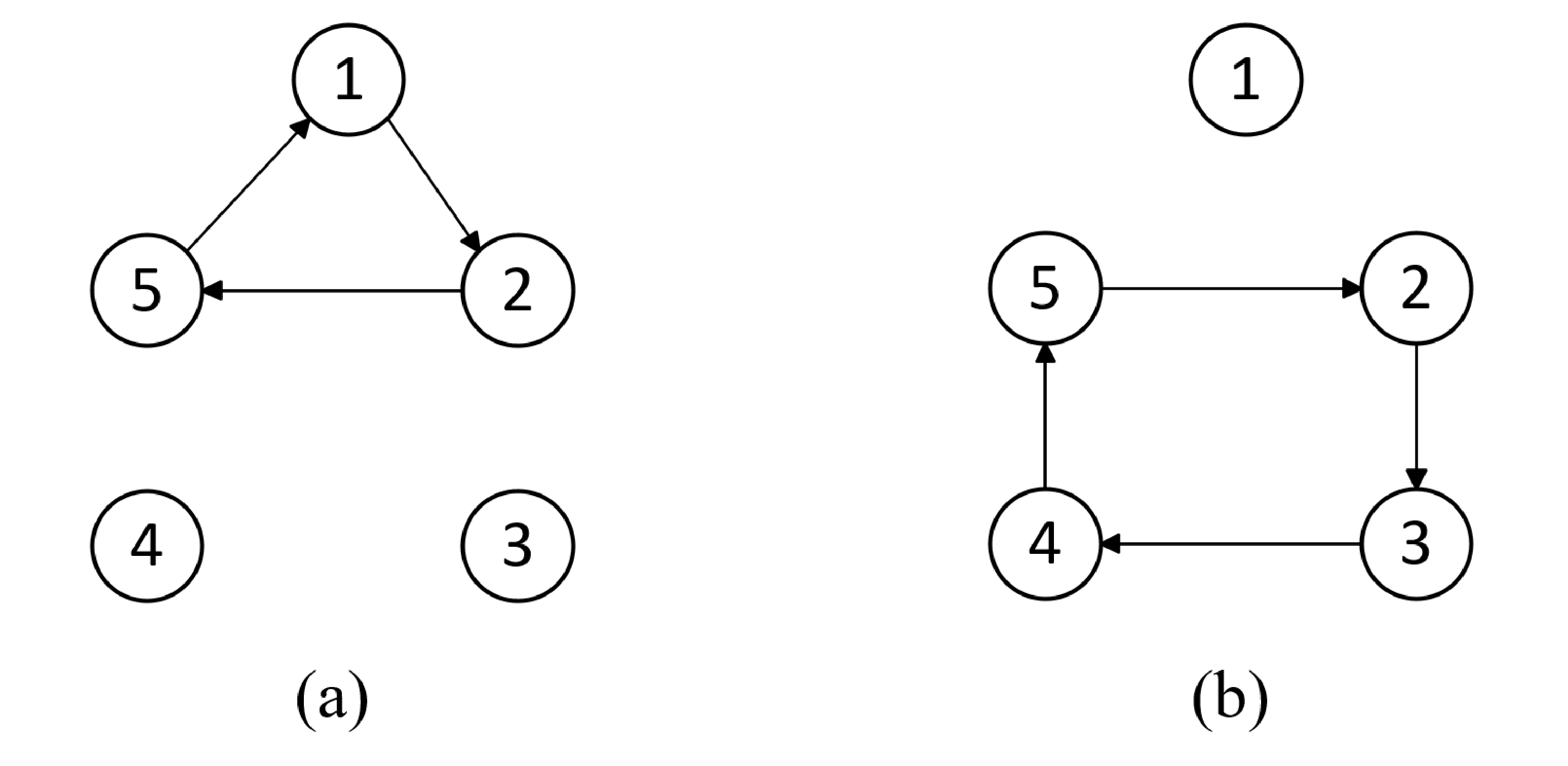}
\caption{The time-varying digraph.} \label{fig8}
\end{figure}

\begin{figure}
\centering
\includegraphics[width=0.4\textwidth]{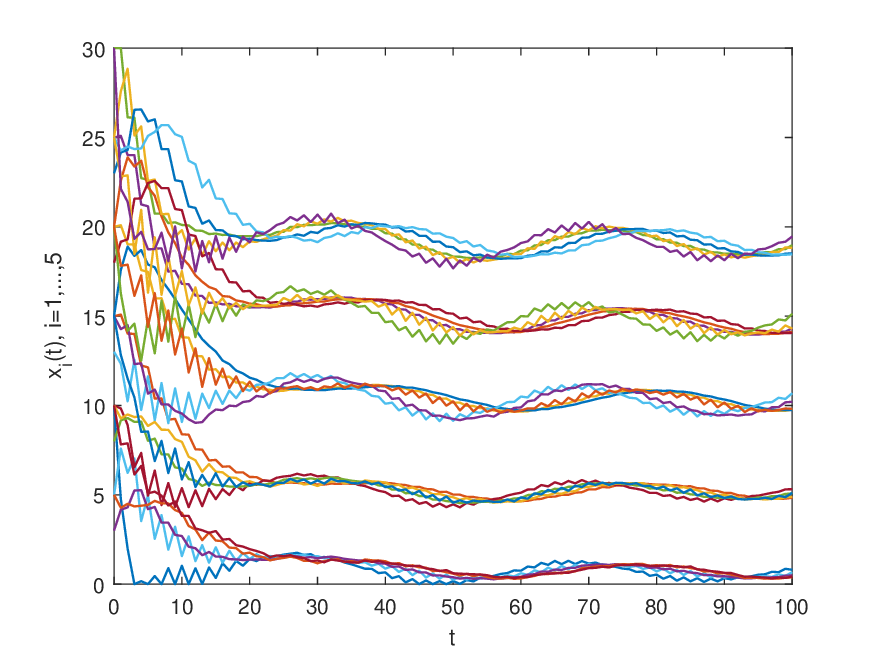}
\caption{The evolutions of $x_i(t)$ under algorithm (\ref{b1}), $i=1,\cdots,5$.} \label{fig9}
\end{figure}

\begin{figure}
\centering
\includegraphics[width=0.4\textwidth]{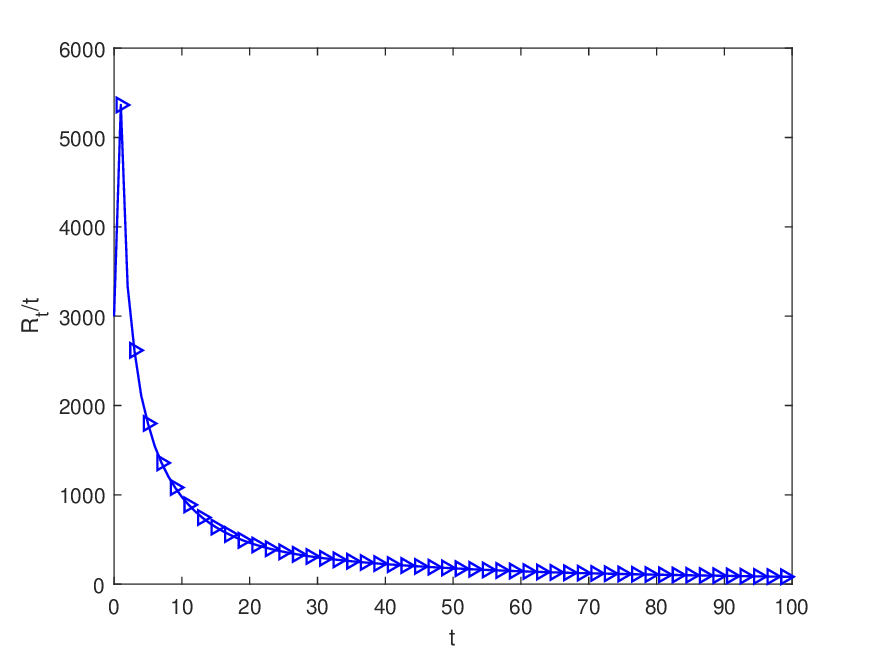}
\caption{The evolution of $\mathcal{R}_{t}/t$ under algorithm (\ref{b1}).} \label{fig10}
\end{figure}

\begin{figure}
\centering
\includegraphics[width=0.4\textwidth]{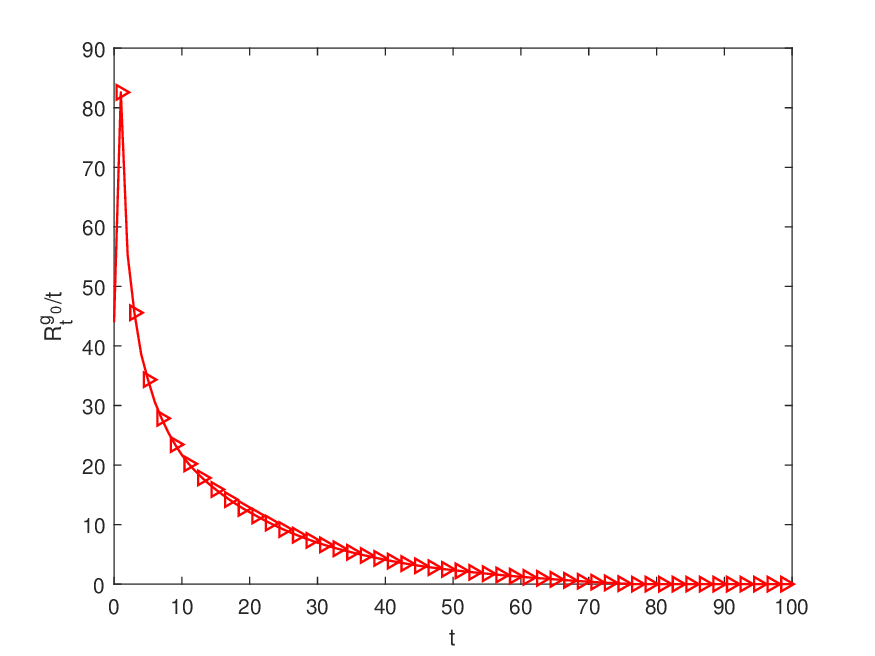}
\caption{The evolution of $\mathcal{R}^{g}_{t}/t$ under algorithm (\ref{b1}).} \label{fig11}
\end{figure}

\textbf{Example 2}~
Now we apply algorithm (\ref{b1}) to addressing an online Nash-Cournot game. In this game, there are five companies that produce the same production, and these companies are indexed by set $\mathcal{V}=\{1,\cdots,5\}$. For any $i\in\mathcal{V}$, let $\textrm{x}_i$ be the quantity produced by company $i$. Due to the influence of some changeable factors such as marginal costs, the production cost
and the demand price may be time-varying \cite{htx35}. For any time $t\in \lfloor T\rfloor$, the production cost and the demand price are given by $p_i^t(\textrm{x}_i)=k_i^t\textrm{x}_i$ and $d_i^t(x)=l_i^t+\theta_i^t-\sum_{j=1}^5\textrm{x}_j$ respectively, where $x=[\textrm{x}_1,\cdots,\textrm{x}_5]^T$, $\theta_i^t$ is a stochastic variable caused by the relation between supply and demand. Then the cost function of  company $i$ follows that $C_i^t(x,\theta_i^t)=p_i^t(\textrm{x}_i)-q_id_i^t(x)$. Moreover, the production constraint of company $i$ is given by $\Omega_i\subset \mathbb{R}$, and the market capacity constraint is given by a coupled inequality constraint $\sum_{i=1}^5g_i^t({x})\leq 0$. In order to maximize benefits, each company aims to minimize its own expected cost function $\mathcal{C}_i^t(x)=\mathbb{E}[C_i^t(x,\theta_i^t)]$. Define $\varphi^t(x)=[\nabla_{\textrm{x}_1}\mathcal{C}_1^t(x),\cdots,\nabla_{\textrm{x}_5}\mathcal{C}_5^t(x)]^T$, $\varphi_i^t(x)=[0,\cdots,0,\nabla_{\textrm{x}_i}\mathcal{C}_i^t(x),0,\cdots,0]^T$, $\Omega=\prod_{i=1}^5\Omega_i$ and $\mathcal{{X}}^t=\left\{{x}\in \Omega\Big|\sum_{i=1}^5g_i^t({x})\leq0\right\}$, by combining game theory and variational inequality theory, we know that seeking the Nash equilibrium of this game is equivalent to finding a $x^*_t\in\mathcal{{X}}^t$ such that
\begin{eqnarray*}
\begin{split}
\langle\varphi^t(x^*_t),y-x^*_t\rangle=\sum_{i=1}^5\langle\varphi_i^t(x^*_t),y-x^*_t\rangle\geq 0,~\forall{y}\in \mathcal{{X}}^t.
\end{split}
\end{eqnarray*}
In this formulation, we set $k_i^t=5\sin\frac{t}{6}$, $l_i^t=5i+45-2.5i\sin\frac{t}{6}$, $\Omega_i=\{\textrm{x}|0\leq \textrm{x}\leq 30\}$, $\theta_i^t\sim \mathcal{N}(0,\frac{1}{2})$, $g_i^t(x)=e_i^Tx+\epsilon_i$, where $e_i$ is a vector with the $i$th entry being one and others being zero, $\epsilon_1=10$, $\epsilon_2=\epsilon_5=15$, $\epsilon_3=\epsilon_4=8$. At each time $t$, Nash equilibria can be computed as $x^*_t=[|\frac{35}{12}\sin\frac{t}{6}|;5+\frac{5}{12}\sin\frac{t}{6};10-\frac{25}{12}\sin\frac{t}{6};15-\frac{55}{12}\sin\frac{t}{6};20-\frac{85}{12}\sin\frac{t}{6}]$. Consider a network of five agents and agents communicate with each other through a time-varying graph shown in Fig. \ref{fig8}, where (a) and (b) are two graphs in a period with $U=2$. The agents solve this problem by running algorithm (\ref{eq8}), where each agent $i$ maintains an estimate on the Nash equilibria denoted by $x_i(t)\in \mathbb{R}^5$ . Here we set
the  time horizon to be $T=100$. The parameters in algorithm (\ref{eq8}) are given by $\zeta=1/\sqrt{T+30}$, $\eta=1/(T+30)^{\frac{1}{3}}$, $\mathcal{D}_\phi({x}, z_i(t))=\|{x}-z_i(t)\|^2$, $x_1(0)=x_5(0)=[10;15;20;25;30]$, $x_2(0)=x_4(0)=[5;10;15;20;25]$, $x_3(0)=[3;8;13;18;23]$. By running algorithm (\ref{b1}) in a single round, Figs. \ref{fig9}, \ref{fig10} and \ref{fig11} show the evolutions of $x_i(t)$, $\frac{\mathcal{R}_{t}}{t}$ and $\frac{\mathcal{R}^{g}_{t}}{t}$, respectively, where $\mathcal{R}_t=\max_{1\leq i\leq5}\mathcal{R}_{i,t}$ and $\mathcal{R}^{g}_{t}=\max_{1\leq i\leq5}\mathcal{R}^{g}_{i,t}$. From Fig. \ref{fig9}, we see that the states of all agents asymptotically approach to $x^*_t$. Furthermore, observing Fig. \ref{fig10} and Fig. \ref{fig11}, one can find that both $\frac{\mathcal{R}_{t}}{t}$ and $\frac{\mathcal{R}^{g}_{t}}{t}$ asymptotically decay to zero. These observations illustrate the validity of Theorem 2.


\section{Conclusions}\label{se5}
In this paper, the problem of online MEPs
with coupled inequality constraints has
been studied. To solve this problem, first, an online distributed algorithm
with accurate gradient information is proposed based on mirror descent algorithms and primal-dual strategies. By running the
algorithm, each agent makes decisions only using the information of local bifunctions and local constraint functions at the previous time,  and the state information received from its immediate neighbors. Dynamic regrets and constraint violations are used to measure the performance of the algorithm. The result shows that if the graph is $U$-strongly
connected, and if the deviation in the solution sequence is within a certain range, then
dynamic regrets and the constraint violations grow sublinearly. Second, considering the case
where the gradients are influenced by noises, we propose an online distributed algorithm involving stochastic gradient information.
We show that under the same conditions as those in the previous case,  if stochastic gradients
satisfy the sub-Gaussian noise condition, then both the dynamic regret and the constraint violation have a high probability sublinear bound. Finally, two simulation examples have been carried out to demonstrate the effectiveness of our theoretical results. How to accelerate the convergence rate of online distributed algorithms for online MEPs is an interesting but
challenging topic, which will be considered in our future work. Some other issues such as time delays and communication bandwidth
constraints will also be considered, which will bring new challenges in the study of online MEPs.

\section{Appendix}\label{se6}

\emph{A.  Proof of Lemma 3}

By the second equation in (\ref{eq8}), using the fact that $\|{a}\|^2\leq \|{b}\|^2$ if ${a}=[b]_+$ for any $a,b\in \mathbb{R}^m$ yields
\begin{eqnarray*}\begin{split}
&\sum_{i=1}^n\Big\|y_i(t+1)\Big\|^2\\
&\leq\sum_{i=1}^n \Big\|\sum_{j=1}^na_{ij}(t)\Big((1-\eta_{t})y_j(t)+\eta_{t}g_i^t(x_i(t))\Big)\Big\|^2\\ &\leq\sum_{i=1}^n\Big(\sum_{j=1}^na_{ij}(t)\Big\|(1-\eta_{t})y_j(t)+\eta_{t}g_i^t(x_i(t))\Big\|\Big)^2\\
&\leq\sum_{i=1}^n\sum_{j=1}^na_{ij}(t)\Big((1-\eta_{t})\|y_j(t)\|^2+\eta_{t}\|g_i^t(x_i(t))\|^2\Big)\\
&\leq(1-\eta_{t})\sum_{i=1}^n\|y_i(t)\|^2+\eta_{t}n\kappa_2^2
\end{split}\end{eqnarray*}
where the second inequality results from triangle inequality, the third one holds by using Jensen's inequality, and the last one results from the facts that $\sum_{j=1}^na_{ij}(t)=\sum_{i=1}^na_{ij}(t)=1$ and (iii) in Assumption 2.
Since $y_i(0)=0$,  there holds that $\sum_{i=1}^n\|y_i(1)\|^2\leq n\kappa_2^2$ and $\sum_{i=1}^n\|y_i(2)\|^2\leq(1-\eta_{1})\sum_{i=1}^n\|y_i(1)\|^2+\eta_{1}n\kappa_2^2\leq n\kappa_2^2$. In a similar fashion, it immediately leads to the validity of the result.

\emph{B.  Proof of Lemma 4}

In Lemma \ref{le2}, letting $\hat{x}={x}_i(t+1)$, ${y}={z}_i(t)$, $w=x^*_t$ and ${s}=\zeta_t\nabla_2 f_i^t(x_i(t),{x}_i(t))+\eta_t\nabla g_i^t(x_i(t)){y}_i(t)$, we have
\begin{eqnarray}\label{eq22}
\begin{split}
&\zeta_t\langle \nabla_2 f_i^t(x_i(t), {x}_i(t)), {x}_i(t+1)-{x}^*_t\rangle\\
&\leq\eta_t\langle\nabla g_i^t(x_i(t)){y}_i(t), {x}^*_t-{x}_i(t+1)\rangle\\
&~~~+\mathcal{D}_\phi(x^*_t, {z}_i(t))-\mathcal{D}_\phi(x^*_t, x_i(t+1))\\
&~~~-\mathcal{D}_\phi(x_i(t+1), {z}_i(t))+\mathcal{D}_\phi(x^*_{t+1}, x_i(t+1))\\
&~~~-\mathcal{D}_\phi(x^*_{t+1}, x_i(t+1))\\
&\leq\eta_t\langle\nabla {g_i^t}(x_i(t)){y}_i(t), {x}^*_t-x_i(t)\rangle\\
&~~~+\eta_t\langle\nabla {g_i^t}(x_i(t)){y}_i(t), {x}_i(t)-x_i(t+1)\rangle\\
&~~~+\mathcal{D}_\phi(x^*_t, {z}_i(t))-\mathcal{D}_\phi(x^*_{t+1}, x_i(t+1))\\
&~~~+\ell\|x^*_{t+1}-x^*_{t}\|\\
&\leq\eta_t({y}_i(t))^T(g_i^t({x}^*_t)-{\color{blue}g_i^t}(x_i(t)))\\
&~~~+{\sqrt{n}h\kappa_2\kappa_3}\eta_t\|x_i(t)-{x}_i(t+1)\|\\
&~~~+\mathcal{D}_\phi(x^*_t, {z}_i(t))-\mathcal{D}_\phi(x^*_{t+1}, x_i(t+1))\\
&~~~+\ell\|x^*_{t+1}-x^*_{t}\|\\
  \end{split}
\end{eqnarray}
where the second inequality holds by using (ii) in Assumption \ref{as4} and the fact that  $\mathcal{D}_\phi(\cdot, \cdot)\geq 0$, the third inequality results from the convexity of $g_{i}^t(\cdot)$, Lemma \ref{le4} and (iii) in Assumption \ref{as2}. By the convexity of $f_i^t(\cdot,\cdot)$ with respect to the second argument, there holds
\begin{eqnarray}\label{eq23}
\begin{split}
 &\sum_{i=1}^n\langle\nabla_2f_i^t(x_i(t), {x}_i(t)), {x}_i(t+1)-{x}_t^*\rangle\\
 &=\sum_{i=1}^n\langle\nabla_2f_i^t(x_i(t), {x}_i(t)), {x}_i(t)-{x}_t^*\rangle\\
 &~~~+\sum_{i=1}^n\langle\nabla_2f_i^t(x_i(t), {x}_i(t)), {x}_i(t+1)-{x}_i(t)\rangle\\
 &\geq \sum_{i=1}^nf_i^t(x_i(t), {x}_i(t))-\sum_{i=1}^nf_i^t(x_i(t), {x}_t^*)\\
 &~~~-\sum_{i=1}^n\kappa_1\|x_i(t+1)-x_i(t)\|\\
  &=-\sum_{i=1}^nf_i^t(x_i(t), {x}_t^*)-\sum_{i=1}^n\kappa_1\|x_i(t+1)-x_i(t)\|\\
  \end{split}
\end{eqnarray}
where the  inequality holds by using (iii) in Assumption \ref{as2}, and the last equality results from (i) in Assumption \ref{as3}. Moreover,
\begin{eqnarray}\label{eq23-1}
\begin{split}
&\eta_t\sum_{i=1}^n(y_i(t))^T(g_i^t(x^*_t)-g_i^t(x_i(t)))\\
&\leq-\eta_t\sum_{i=1}^n(y_i(t))^Tg_i^t(x_i(t))+\eta_t\sum_{i=1}^n(\bar{y}(t))^Tg_i^t(x^*_t)\\
&~~~+\eta_t\sum_{i=1}^n(y_i(t)-\bar{y}(t))^Tg_i^t(x^*_t)\\
&\leq-\eta_t\sum_{i=1}^n(y_i(t))^Tg_i^t(x_i(t))+\kappa_2\eta_t\sum_{i=1}^n\|y_i(t)-\bar{y}(t)\|\\
  \end{split}
\end{eqnarray}
where the last step holds by using $\sum_{i=1}^n$ $(\bar{{y}}(t))^Tg_i^t(x^*_t)\leq0$ and (iii) in Assumption \ref{as2}. By the balance condition of $\mathcal{G}(t)$ and the convexity of $\mathcal{D}_\phi(\cdot,\cdot)$ with respect to the second argument, we know that
$\sum_{i=1}^n\mathcal{D}_\phi(x^*_t, {z}_i(t))\leq\sum_{i=1}^n\mathcal{D}_\phi(x^*_t, x_i(t))$.
Together with (\ref{eq22})-(\ref{eq23-1}) and the fact that $\zeta_t\leq \eta_t$, it immediately results in that
\begin{eqnarray}\label{eq23-2}
\begin{split}
&-\zeta_t\sum_{i=1}^nf_i^t(x_i(t), {x}_t^*)\\
&\leq-\eta_t\sum_{i=1}^n(y_i(t))^Tg_i^t(x_i(t))+\kappa_2\eta_t\sum_{i=1}^n\|y_i(t)-\bar{y}(t)\|\\
&~~~+\rho\eta_t\sum_{i=1}^n\|x_i(t)-x_i(t+1)\|+n\ell\|x^*_{t+1}-x^*_{t}\|\\
&~~~+\sum_{i=1}^n\Big(\mathcal{D}_\phi(x^*_t, x_i(t))-\mathcal{D}_\phi(x^*_{t+1}, x_i(t+1))\Big).
  \end{split}
\end{eqnarray}
Using (ii) in Assumption \ref{as3}, for any $i,j\in\mathcal{V}$, there holds
 \begin{eqnarray*}\begin{split}
 &f^t(x_i(t), x^*_t)=\sum_{j=1}^nf_j^t(x_i(t), x^*_t)\\
 &=\sum_{j=1}^nf_j^t(x_i(t), x^*_t)-\sum_{j=1}^nf_j^t(x_j(t), x^*_t)+\sum_{j=1}^nf_j^t(x_j(t), x^*_t)\\
 \end{split}\end{eqnarray*}
  \begin{eqnarray}\label{eq27}\begin{split}
 &\geq -nL\|x_i(t)-x_j(t)\|+\sum_{j=1}^nf_j^t(x_j(t), x^*_t)\\
  &\geq -2nL\|x_i(t)-\bar{x}(t)\|+\sum_{j=1}^nf_j^t(x_j(t), x^*_t).\\
\end{split}\end{eqnarray}
Combining (\ref{eq23-2}) and (\ref{eq27}) immediately implies (\ref{eq24}).

\emph{C.  Proof of Lemma 5}

(i) In Lemma \ref{le2}, letting $\hat{x}=x_i(t+1)$, $w=y=z_i(t)$ and ${s}=\zeta_t\nabla_2 f_i^t(x_i(t),{x}_i(t))+\eta_t\nabla g_i^t(x_i(t)){y}_i(t)$, combining the fact that $\mathcal{D}_\phi(x,y)\geq\frac{\mu}{2}\|{x}-{y}\|^2$ results in that
\begin{eqnarray*}\label{eq11}
\begin{split}
&\mu\|{x}_i(t+1)-{z}_i(t)\|^2\\
&\leq \mathcal{D}_\phi(x_i(t+1),z_i(t))+\mathcal{D}_\phi(z_i(t),x_i(t+1))\\
&\leq \big\langle \zeta_t\nabla_2 f_i^t(x_i(t), x_i(t))+\eta_t\nabla g_i^t(x_i(t))y_i(t),\\
&~~~ {z}_i(t)-{x}_i(t+1)\big\rangle\\
&\leq\eta_t \Big(\|\nabla_2 f_i^t(x_i(t), x_i(t))\|+\\
&~~~\|y_i(t)\|\sum_{k=1}^h\|\nabla g_{ik}^t(x_i(t))\|\Big)\|x_i(t+1)-z_i(t)\|\\
&\leq ({\sqrt{n}h\kappa_2\kappa_3}+\kappa_1)\eta_t\|x_i(t+1)-z_i(t)\|
  \end{split}
\end{eqnarray*}
where the last inequality results by Lemma \ref{le4} and Assumption \ref{as2}. It immediately implies that
\begin{eqnarray}\label{eq11-1}
\begin{split}
\|{x}_i(t+1)-{z}_i(t)\|\leq \frac{\sqrt{n}h\kappa_2\kappa_3+\kappa_1}{\mu}\eta_t.
  \end{split}
\end{eqnarray}
Denote $e_i(t)=x_i(t+1)-z_i(t)$, one has $\|e_i(t)\|\leq \frac{\sqrt{n}h\kappa_2\kappa_3+\kappa_1}{\mu}\eta_t$ and
\begin{eqnarray}\label{9}\begin{split}
x_i(t+1)=\sum_{j\in\mathcal{N}_i(t)}a_{ij}(t)x_j(t)+e_i(t).
\end{split}\end{eqnarray}
Let $\tilde{x}_r(t)\in\mathbb{R}^{n}$ and $\tilde{e}_r(t)\in\mathbb{R}^{n}$ denote the stack of the $r^{th}$ terms of all $x_i(t)$ and the stack of the $r^{th}$ term of all $e_i(t)$ respectively, where $r\in\{1,\cdots, m\}$. Based on (\ref{9}), we have
\begin{eqnarray*}\begin{split}
\tilde{x}_r(t+1)=A(t)\tilde{x}_r(t)+\tilde{e}_r(t)
\end{split}\end{eqnarray*}
which implies that
\begin{eqnarray}\label{10}\begin{split}
\tilde{x}_r(t)=\Phi(t,0)\tilde{x}_r(0)+\sum_{s=1}^t\Phi(t,s)\tilde{e}_r(s-1)
\end{split}\end{eqnarray}
where $\Phi(t,s)$ is defined in (\ref{eq0}). Note that $\Phi(t,s)$ is a doubly
stochastic matrix, then we have
\begin{eqnarray}\label{11}\begin{split}
1_n^T\tilde{x}_r(t)=1_n^T\tilde{x}_r(0)+\sum_{s=1}^t1_n^T\tilde{e}_r(s-1).
\end{split}\end{eqnarray}
Combining (\ref{10}) and (\ref{11}) results in that for any $i\in\mathcal{V}$,
\begin{eqnarray*}\begin{split}
&\Big|[\tilde{x}_r(t)]_{i\cdot}-\frac{1}{n}1_n^T{\tilde{x}}_r(t)\Big|\\
&\leq\left|\left([{\Phi}(t,0)]_{i\cdot}-\frac{1}{n}{1}_n^T\right)\tilde{x}_r(0)\right|+\\
&~~~~\sum_{s=1}^t\left|\left([{\Phi}(t,s)]_{i\cdot}-\frac{1}{n}{1}_n^T\right)\tilde{e}_r(s-1)\right|\\
&\leq\max_{1\leq j\leq n}\left|[{\Phi}(t,0)]_{ij}-\frac{1}{n}\right|\left\|\tilde{x}_r(0)\right\|_1+\\
&~~~~\sum_{s=1}^t\max_{1\leq j\leq n}\left|[{\Phi}(t,s)]_{ij}-\frac{1}{n}\right|\left\|\tilde{e}_r(s-1)\right\|_1.
\end{split}\end{eqnarray*}
Using Lemma \ref{le1} and the fact that  $\|\tilde{x}_r(0)\|_1\leq n\kappa$ yields
\begin{eqnarray*}\begin{split}
&\Big|[\tilde{x}_r(t)]_{i\cdot}-\frac{1}{n}1_n^T{\tilde{x}}_r(t)\Big|\\
&\leq n\kappa\mathcal{C}\lambda^t+\frac{{n}^{\frac{3}{2}}h\kappa_2\kappa_3\mathcal{C}+n\kappa_1\mathcal{C}}{\mu\lambda}\sum_{s=0}^t\lambda^{t-s}\eta_{s}.
\end{split}\end{eqnarray*}
It immediately implies the validity of (i).

(ii) By the facts that $\eta_{t}\leq1$ and $0<\lambda<1$,  we have
\begin{eqnarray}\label{12}\begin{split}
&\|x_i(t)-\bar{x}(t)\|^2\\
&\leq \Big(mn^2\kappa^2\mathcal{C}^2+\frac{2mn^{\frac{5}{2}}h\kappa\kappa_2\kappa_3\mathcal{C}^2+2mn^2\kappa\kappa_1\mathcal{C}^2}{\mu\lambda(1-\lambda)}\Big)\lambda^t\\
&~~~~+\frac{m({n}^{\frac{3}{2}}h\kappa_2\kappa_3\mathcal{C}+n\kappa_1\mathcal{C})^2}{\mu^2\lambda^2}\Big(\sum_{s=0}^t\lambda^{t-s}\eta_{s}\Big)^2.
\end{split}\end{eqnarray}
Using Cauchy-Schwarz inequality yields
\begin{eqnarray*}\begin{split}
\Big(\sum_{s=0}^t\lambda^{t-s}\eta_{s}\Big)^2 &\leq\Big(\sum_{s=0}^t\lambda^{t-s}\Big)\Big(\sum_{s=0}^t\lambda^{t-s}\eta_{s}^2\Big)\\
&\leq\frac{1}{1-\lambda}\sum_{s=0}^t\lambda^{t-s}\eta_{s}^2.
\end{split}\end{eqnarray*}
Together with (\ref{12}), it immediately leads to the validity of (ii).

(iii) Note that for any $i,j\in \mathcal{V}$,
\begin{eqnarray}\label{eq11-2}
\begin{split}
&\|{x}_i(t)-x_i(t+1)\|\\
&\leq \|{x}_i(t)-z_i(t)\|+\|z_i(t)-x_i(t+1)\|\\
&\leq \|{x}_i(t)-x_j(t)\|+\|z_i(t)-x_i(t+1)\|\\
&\leq 2\|{x}_i(t)-\bar{x}(t)\|+\|z_i(t)-x_i(t+1)\|.
  \end{split}
\end{eqnarray}
Using (\ref{eq11-1}) and (i) immediately yields (iii).

(iv) By the second equation in (\ref{eq8}), there holds that
\begin{eqnarray}\label{eq9}
\begin{split}
 &\left\|{y}_i(t+1)-\sum_{j\in\mathcal{N}_i(t)}a_{ij}(t){y}_{j}(t)\right\|\\
  &\leq\Big\|-\eta_t\sum_{j\in\mathcal{N}_i(t)}a_{ij}(t){y}_{j}(t)+\eta_tg_i^t(x_i(t)) \Big\| \\
   &\leq\eta_t\Big(\sum_{j\in\mathcal{N}_i(t)}a_{ij}(t)\|y_{j}(t)\|+\|g_i^t(x_i(t))\|\Big)\\
  &\leq(\sqrt{n}+1)\kappa_2\eta_t
  \end{split}
\end{eqnarray}
where the first inequality holds by using the fact that $\|[a]_+-[b]_+\|\leq \|a-b\|$ for any $a,b\in \mathbb{R}^m$, the second inequality holds by using the triangle inequality, and the last one results from Lemma \ref{le4} and (iii) in Assumption \ref{as2}. Now we let $\xi_i(t)=y_i(t+1)-\sum_{j\in\mathcal{N}_i(t)}a_{ij}(t)y_{j}(t)$, then,
\begin{eqnarray}\label{eq17}
\begin{split}
 &y_i(t+1)=\sum_{j\in\mathcal{N}_i(t)}a_{ij}(t)y_{j}(t)+\xi_i(t).
  \end{split}
\end{eqnarray}
It can be found that equation (\ref{eq17}) is similar to equation (\ref{9}). Applying same analysis approaches as used in (i) to (\ref{eq17}), and combining the fact that $y_i(0)=0$ for any $i\in\mathcal{V}$, we can verify the validity of (iv).

(v) Based on the result in (iv), using the same method as used in the proof of (ii), one can obtain (v).

\emph{D.  Proof of Lemma 6}

By the second equation in (\ref{eq8}), for any $y\in\mathbb{R}^h_+$, there is
\begin{eqnarray}\label{24}\begin{split}
&\sum_{i=1}^n\Big\|y_i(t+1)-y\Big\|^2\\
&=\sum_{i=1}^n\Big\|\Big[(1-\eta_{t})\sum_{j=1}^na_{ij}(t)y_j(t)+\eta_{t}g_i^t(x_i(t))\Big]_+-y\Big\|^2\\
&\leq\sum_{i=1}^n\Big\|(1-\eta_{t})\sum_{j=1}^na_{ij}(y_j(t)-y_i(t))+(y_i(t)-y)\\
&~~~~+\eta_{t}(g_i^t(x_i(t))-y_i(t))\Big\|^2\\
&\leq\sum_{i=1}^n\Big(\sum_{j=1}^na_{ij}(t)\|y_j(t)-y_i(t)\|^2+\|y_i(t)-y\|^2\\
&~~~~+\eta_{t}^2\|g_i^t(x_i(t))-y_i(t)\|^2\\
&~~~~+2(1-\eta_{t})\sum_{j=1}^na_{ij}(t)\langle y_j(t)-y_i(t),y_i(t)-y\rangle\\
&~~~~+2\eta_{t}\sum_{j=1}^na_{ij}(t)\|y_j(t)-y_i(t)\|\|g_i^t(x_i(t))-y_i(t)\|\\
&~~~~+2\eta_{t}\langle y_i(t)-y,g_i^t(x_i(t))-y_i(t)\rangle\Big)
\end{split}\end{eqnarray}
where the first inequality holds by using the fact that for any $a,b\in \mathbb{R}^m$,  $\|[a]_+-[b]_+\|^2\leq \|a-b\|^2$.
For the first term on the right-hand side of (\ref{24}), we have
\begin{eqnarray}\label{25}\begin{split}
&\sum_{i=1}^n\sum_{j=1}^na_{ij}(t)\|y_j(t)-y_i(t)\|^2\\
&=\sum_{i=1}^n\sum_{j=1}^na_{ij}(t)\|y_j(t)-\bar{y}(t)-(y_i(t)-\bar{y}(t))\|^2\\
&\leq4\sum_{i=1}^n\|y_i(t)-\bar{y}(t)\|^2.
\end{split}\end{eqnarray}
Furthermore, the fourth term on the right-hand side of (\ref{24}) can be manipulated as
 \begin{eqnarray}\label{26}\begin{split}
&\sum_{i=1}^n\sum_{j=1}^na_{ij}(t)\langle y_j(t)-y_i(t),y_i(t)-y\rangle\\
&=\sum_{i=1}^n\sum_{j=1}^na_{ij}(t)\langle y_j(t)-y_i(t),y_i(t)-\bar{y}(t)\rangle\\
\end{split}\end{eqnarray}
\begin{eqnarray*}\begin{split}
&\leq \frac{1}{2}\sum_{i=1}^n\sum_{j=1}^na_{ij}(t)\Big(\|y_j(t)-y_i(t)\|^2\\
&~~~~+\|y_i(t)-\bar{y}(t)\|^2\Big)\\
&\leq \frac{5}{2}\sum_{i=1}^n\|y_i(t)-\bar{y}(t)\|^2\\
\end{split}\end{eqnarray*}
where the equation results by using the fact that $\sum_{i=1}^n\sum_{j=1}^na_{ij}(t)\langle y_j(t)-y_i(t),y\rangle=0$ for any $y\in\mathbb{R}^h_+$.  Noting that the last term of (\ref{24}) can be manipulated as
\begin{eqnarray}\label{27}\begin{split}
&\langle  y_i(t)-y,g_i^t(x_i(t))-y_i(t)\rangle\\
&\leq (g_i^t(x_i(t)))^Ty_i(t)- (g_i^t(x_i(t)))^Ty\\
&~~~~+y^Ty_i(t)-\|y_i(t)\|^2\\
&\leq (g_i^t(x_i(t)))^Ty_i(t)-(g_i^t(x_i(t)))^Ty+\frac{\|y\|^2}{2}
\end{split}\end{eqnarray}
where the inequality results from the fact that $y^Ty_i(t)\leq\frac{1}{2}(\|y\|^2+\|y_i(t)\|^2)$. By (\ref{25}), we can conclude that $\sum_{i=1}^n\sum_{j=1}^na_{ij}(t)\|y_j(t)-y_i(t)\|\leq 2\sum_{i=1}^n\|y_i(t)-\bar{y}(t)\|$.
By Lemma \ref{le4}, we know that $\|g_i^t(x_i(t))-y_i(t)\|\leq\kappa_2+\sqrt{n}\kappa_2$ and $\|g_i^t(x_i(t))-y_i(t)\|^2\leq2(\|g_i^t(x_i(t))\|^2+\|y_i(t)\|^2)=2\kappa_2^2+2n\kappa_2^2$.
Together with (\ref{25})-(\ref{27}),   it follows from (\ref{24}) that
\begin{eqnarray}\label{28}\begin{split}
&\eta_t\sum_{i=1}^n\Big(g_i^t(x_i(t))\Big)^Ty-\eta_t\sum_{i=1}^n\Big(g_i^t(x_i(t))\Big)^Ty_i(t)\\
&\leq \frac{1}{2}\sum_{i=1}^n\Big(\|y_i(t)-y\|^2-\|y_i(t+1)-y\|^2\Big)\\
&~~~~+\eta_t^2(n\kappa_2^2+n^2\kappa_2^2)+\frac{9}{2}\sum_{i=1}^n\|y_i(t)-\bar{y}(t)\|^2\\
&~~~~+2\eta_t(\kappa_2+\sqrt{n}\kappa_2)\sum_{i=1}^n\|y_i(t)-\bar{y}(t)\|+\eta_t\frac{n\|y\|^2}{2}.
\end{split}\end{eqnarray}
Letting $y=0$ immediately implies (\ref{8-1}).

\emph{E.  Proof of Theorem 1}

By (\ref{eq24}) in Lemma \ref{le5} and (\ref{8-1}) in Lemma \ref{le7}, we have
\begin{eqnarray}\label{eq25-1}
\begin{split}
&- f^t(x_i(t), x^*_t)\\
&\leq \frac{1}{2\zeta_t}\sum_{i=1}^n\Big(\|y_i(t)\|^2-\|y_i(t+1)\|^2\Big)\\
&~~~+\frac{\eta_t^2}{\zeta_T}(n\kappa_2^2+n^2\kappa_2^2)+\frac{9}{2\zeta_T}\sum_{i=1}^n\|y_i(t)-\bar{y}(t)\|^2\\
&~~~+\frac{\eta_t}{\zeta_T}(3\kappa_2+2\sqrt{n}\kappa_2)\sum_{i=1}^n\|y_i(t)-\bar{y}(t)\|\\
&~~~+\frac{\eta_t}{\zeta_T}\rho\sum_{i=1}^n\|x_i(t)-x_i(t+1)\|\\
&~~~+\frac{1}{\zeta_t}\sum_{i=1}^n\Big(\mathcal{D}_\phi(x^*_t, x_i(t))-\mathcal{D}_\phi(x^*_{t+1}, x_i(t+1))\Big)\\
&~~~+\frac{1}{\zeta_T}n\ell\|x^*_{t+1}-x^*_{t}\|+\frac{\eta_t}{\zeta_T}2nL\|x_i(t)-\bar{x}(t)\|.
\end{split}
\end{eqnarray}
Note that for any $i\in\mathcal{V}$, there holds
\begin{eqnarray}\label{12-2}\begin{split}
&\sum_{t=0}^T\frac{1}{\zeta_t}\left(\mathcal{D}_\phi(x^*_t, x_i(t))-\mathcal{D}_\phi(x^*_{t+1}, x_i(t+1))\right)\\
&=\frac{1}{\zeta_0}\mathcal{D}_\phi(x^*_0, x_i(0))-\frac{1}{\zeta_T}\mathcal{D}_\phi(x^*_{t+1}, x_i(t+1))\\
&~~~~+\sum_{t=1}^T\left(\frac{1}{\zeta_t}-\frac{1}{\zeta_{t-1}}\right)\mathcal{D}_\phi(x^*_t, x_i(t))\\
&\leq\frac{4K\kappa^2}{\zeta_0}+4K\kappa^2\sum_{t=1}^T\left(\frac{1}{\zeta_t}-\frac{1}{\zeta_{t-1}}\right)\\
&= \frac{4K\kappa^2}{\zeta_T}
\end{split}\end{eqnarray}
where the inequality holds by using (ii) in Assumption \ref{as2} and (iii) in Assumption \ref{as4}.
Similarly, there also holds
\begin{eqnarray}\label{12-3}\begin{split}
&\sum_{t=0}^T\frac{1}{\zeta_t}\left(\|y_i(t)\|^2-\|y_i(t+1)\|^2\right)\leq \frac{n\kappa^2}{\zeta_T}.
\end{split}\end{eqnarray}
Summing over $t=0,\cdots, T$ at both sides of (\ref{eq25-1}), and  using (\ref{12-2}), (\ref{12-3}) and Lemma \ref{le6}, we have
\begin{eqnarray}\label{eq26}
\begin{split}
&- \sum_{t=0}^Tf^t(x_i(t), x^*_t)\\
&\leq \frac{n\kappa_2^2+n^2\kappa_2^2}{\zeta_T} \sum_{t=0}^T\eta_t^2+\frac{9n\rho_7}{2\zeta_T}\sum_{t=0}^T\sum_{s=0}^{t}\lambda^{t-s}\eta_{s}^2\\
&~~~+\frac{n^2\kappa^2}{2\zeta_T}+\frac{(3\kappa_2+2\sqrt{n}\kappa_2)n\rho_6}{\zeta_T}\sum_{t=0}^T\sum_{s=0}^{t}\lambda^{t-s}\eta_{s}^2\\
&~~~+\frac{2n\rho\rho_1}{\zeta_T}\sum_{t=0}^T\lambda^{t}+\frac{2n\rho\rho_2}{\zeta_T}\sum_{t=0}^T\sum_{s=0}^{t}\lambda^{t-s}\eta_{s}^2\\
&~~~+\frac{n\rho\rho_5}{\zeta_T}\sum_{t=0}^T\eta_t^2+\frac{n\ell}{\zeta_T}\sum_{t=0}^T\|x^*_{t+1}-x^*_{t}\|+\frac{4nK\kappa^2}{\zeta_T}\\
&~~~+\frac{2nL\rho_1}{\zeta_T}\sum_{t=0}^T\lambda^{t}+\frac{2nL\rho_2}{\zeta_T}\sum_{t=0}^T\sum_{s=0}^{t}\lambda^{t-s}\eta_{s}^2.
\end{split}
\end{eqnarray}
It is easy to verify that $\sum_{t=0}^T\lambda^t\leq \frac{1}{1-\lambda}$ and $\sum_{t=0}^T \eta_{t}^2\leq \mathcal{O}(T^{1-2b})$. Moreover,
\begin{eqnarray*}\begin{split}
\sum_{t=0}^T\sum_{s=0}^t\lambda^{t-s}\eta_{s}^2&=\sum_{s=0}^T\sum_{t=s}^T\lambda^{t-s}\eta_{s}^2\\
&\leq \Big(\sum_{s=0}^T\eta_{s}^2\Big)\Big(\sum_{t=0}^T\lambda^{t}\Big)\\
& \leq \mathcal{O}(T^{1-2b}).
\end{split}\end{eqnarray*}
Then (\ref{eq26}) immediately implies (\ref{7-1}).
Now combining (\ref{eq24}) and (\ref{28})  yields that
\begin{eqnarray}\label{30}\begin{split}
&\sum_{i=1}^n\Big(g_i^t(x_i(t))\Big)^Ty-\frac{n\|y\|^2}{2}\\
&\leq \frac{1}{2\eta_t}\sum_{i=1}^n\Big(\|y_i(t)-y\|^2-\|y_i(t+1)-y\|^2\Big)\\
&~~~+\eta_t(\kappa_2^2+n\kappa_2^2)+\frac{9}{2\eta_t}\sum_{i=1}^n\|y_i(t)-\bar{y}(t)\|^2\\
\end{split}\end{eqnarray}
\begin{eqnarray*}\begin{split}
&~~~+(3\kappa_2+2\sqrt{n}\kappa_2)\sum_{i=1}^n\|y_i(t)-\bar{y}(t)\|\\
&~~~+\rho\sum_{i=1}^n\|x_i(t)-x_i(t+1)\|+\frac{\zeta_t}{\eta_t} f^t(x_i(t), x^*_t) \\
&~~~+\sum_{i=1}^n\frac{1}{\eta_t}\Big(\mathcal{D}_\phi(x^*_t, x_i(t))-\mathcal{D}_\phi(x^*_{t+1}, x_i(t+1))\Big)\\
&~~~+\frac{1}{\eta_t}n\ell\|x^*_{t+1}-x^*_{t}\|+2nL\|x_i(t)-\bar{x}(t)\|.
\end{split}\end{eqnarray*}
It is noticed that for any $i\in\mathcal{V}$,
\begin{eqnarray}\label{31}\begin{split}
&\sum_{t=0}^T\frac{1}{\eta_t}\left(\|y_i(t)-y\|^2-\|y_i(t+1)-y\|^2\right)\\
&=\frac{1}{\eta_0}\|y_i(0)-y\|^2-\frac{1}{\eta_T}\|y_i(T+1)-y\|^2\\
&~~~~+\sum_{t=1}^T\left(\frac{1}{\eta_t}-\frac{1}{\eta_{t-1}}\right)\|y_i(t)-y\|^2\\
&\leq\frac{2(n\kappa_2^2+\|y\|^2)}{\eta_0}+2(n\kappa_2^2+\|y\|^2)\sum_{t=1}^T\left(\frac{1}{\eta_t}-\frac{1}{\eta_{t-1}}\right)\\
&= \frac{2(n\kappa_2^2+\|y\|^2)}{\eta_T}.
\end{split}\end{eqnarray}
Using (i) and (ii) in Assumption \ref{as3} yields
\begin{eqnarray}\label{32}\begin{split}
f^t(x_i(t), x^*_t)&=f^t(x_i(t), x^*_t)-f^t( x^*_t, x^*_t)\\
&\leq nL \|x_i(t)-x^*_t\|\leq 2nL\kappa.
\end{split}\end{eqnarray}
By (\ref{30})-(\ref{32}), with reference to the proof of (\ref{7-1}), we can conclude that
\begin{eqnarray}\label{33}\begin{split}
&\sum_{t=0}^T\sum_{i=1}^n\Big(g_i^t(x_i(t))\Big)^Ty-\Big(T+\frac{2}{\eta_T}\Big)\frac{n\|y\|^2}{2}\\
&\leq \mathcal{O}\Big({T}^{1+b-a}+T^b\Theta_T\Big).
\end{split}\end{eqnarray}
Letting $y=\frac{\big[\sum_{t=0}^T\sum_{i=1}^ng_i^t(x_i(t))\big]_+}{n(T+\frac{2}{\eta_T})}$, there holds that
\begin{eqnarray*}\begin{split}
&\Big(\sum_{t=0}^T\sum_{i=1}^ng_i^t(x_i(t))\Big)^Ty-\Big(T+\frac{2}{\eta_T}\Big)\frac{n\|y\|^2}{2}\\
&=\frac{\big[\sum_{t=0}^T\sum_{i=1}^ng_i^t(x_i(t))\big]_+^2}{2n(T+\frac{2}{\eta_T})}.
\end{split}\end{eqnarray*}
Together with (\ref{33}), it follows that
\begin{eqnarray}\label{7-3}\begin{split}
&\left\|\left[\sum_{t=0}^T\sum_{i=1}^ng_i^t(x_i(t))\right]_+\right\|\\
&\leq \mathcal{O}\Big(\sqrt{{T}^{2+b-a}+T^{1+b}\Theta_T}\Big).
\end{split}\end{eqnarray}
By (iii) in Assumption \ref{as2} and (i) in Lemma \ref{le6}, we have
\begin{eqnarray}\label{7-4}\begin{split}
&\left\|\left[\sum_{t=0}^T\sum_{j=1}^ng_j^t(x_i(t))\right]_+\right\|-\left\|\left[\sum_{t=0}^T\sum_{j=1}^ng_j^t(x_j(t))\right]_+\right\|\\
&\leq h\kappa_3\sum_{t=0}^T\sum_{j=1}^n\|x_j(t)-x_i(t)\|\\
&\leq 2h\kappa_3\sum_{t=0}^T\sum_{j=1}^n\|x_j(t)-\bar{x}(t)\|\\
&\leq 2nh\kappa_3\rho_1\sum_{t=0}^T\lambda^{t}+2nh\kappa_3\rho_2\sum_{t=0}^T\sum_{s=0}^{t}\lambda^{t-s}\eta_{s}\\
&\leq \mathcal{O}\Big(T^{1-b}\Big).
\end{split}\end{eqnarray}
Combining (\ref{7-3}) and (\ref{7-4}) immediately implies (\ref{7-2}). This completes the proof of Theorem \ref{TH1}.

\emph{F.  Proof of Lemma 7}

Before presenting the proof of Lemma \ref{le9-1},  a lemma benefit for analyzing high probability bounds of stochastic sequences are presented.
\begin{lemma}\label{le8}\cite{htx34}
Given a random sequence $\theta_t$ and a sequence $\vartheta_t$ for $t=0,\cdots,T$, if  $\mathbb{E}[\exp(\theta_t)|\mathcal{F}_t]\leq \exp(\vartheta_t)$, then for any $\nu\in(0,1)$, $\sum_{t=0}^T\theta_t\leq\sum_{t=0}^T\vartheta_t+\ln \frac{1}{\nu}$ with probability at least $1-\nu$.
\end{lemma}

Now we provide the proof of Lemma \ref{le9-1}.

\textbf{Proof of Lemma 7.} Using exponential inequality $\exp(a)\leq \exp(a^2)+a$ for any $a\in\mathbb{R}$ and taking the conditional expectation yields
\begin{eqnarray}\label{13-1}\begin{split}
&\mathbb{E}\Big[\exp\Big(\frac{\eta }{2\sqrt n\kappa\sigma_1}\sum_{i=1}^n\Big\langle b_i^t-\nabla_2f_i^t(x_i(t),x_i(t)),\\
& x^*_t-x_i(t)\Big\rangle\Big)\Big|\mathcal{F}_t\Big]\\
&\leq\mathbb{E}\Big[\exp\Big(\eta^2\sum_{i=1}^n\frac{\|b_i^t-\nabla_2f_i^t(x_i(t),x_i(t))\|^2}{\sigma_1^2}\Big)\Big|\mathcal{F}_t\Big]\\
&\leq\Big(\prod_{i=1}^n\mathbb{E}\Big[\exp\Big(\frac{\|b_i^t-\nabla_2f_i^t(x_i(t),x_i(t))\|^2}{\sigma_1^2}\Big)\Big|\mathcal{F}_t\Big]\Big)^{\eta^2}\\
&\leq\exp\big(n\eta^2\big)
\end{split}\end{eqnarray}
where the first inequality holds due to the fact that $\|x^*_t-x_i(t+1)\|^2\leq4\kappa^2$,  the second inequality holds by using Jensen's inequality, and the last one is true due to (ii) in Assumption \ref{as5}. Applying (ii) in Lemma \ref{le8} to (\ref{13-1}) immediately implies (\ref{13}). Using the same methods, one can verify the validity of (\ref{13*}).

\emph{G.  Proof of Lemma 8}

In Lemma \ref{le2}, letting $\hat{x}=x_i(t+1)$, $w=y=z_i(t)$ and ${s}=\zeta b_i^t+\eta c_i^t{y}_i(t)$, combining the fact that $\mathcal{D}_\phi(x,y)\geq\frac{\mu}{2}\|{x}-{y}\|^2$ results in that
\begin{eqnarray*}
\begin{split}
&\mu\|{x}_i(t+1)-{z}_i(t)\|^2\\
&\leq \mathcal{D}_\phi(x_i(t+1),z_i(t))+\mathcal{D}_\phi(z_i(t),x_i(t+1))\\
&\leq \big\langle \zeta b_i^t+\eta c_i^t{y}_i(t),{z}_i(t)-{x}_i(t+1)\big\rangle\\
&\leq\eta  \Big(\|\nabla_2 f_i^t(x_i(t), x_i(t))\|+\\
&~~~\|y_i(t)\|\sum_{k=1}^h\|\nabla g_{ik}^t(x_i(t))\|\Big)\|x_i(t+1)-z_i(t)\|\\
&~~~+\eta \|b_i^t-\nabla_2f_i^t(x_i(t),x_i(t))\|\|x_i(t+1)-z_i(t)\|\\
&~~~+\eta \|y_i(t)\|\|c_i^t-\nabla g_i^t(x_i(t))\|_F\|x_i(t+1)-z_i(t)\|\\
&\leq ({\sqrt{n}h\kappa_2\kappa_3}+\kappa_1)\eta \|x_i(t+1)-z_i(t)\|\\
&~~~+\eta \|b_i^t-\nabla_2f_i^t(x_i(t),x_i(t))\|\|x_i(t+1)-z_i(t)\|\\
&~~~+\sqrt{n}\kappa_2\eta \|c_i^t-\nabla g_i^t(x_i(t))\|\|x_i(t+1)-z_i(t)\|\\
  \end{split}
\end{eqnarray*}
where the last inequality results by Lemma \ref{le4} and Assumption \ref{as2}. It immediately implies that
\begin{eqnarray}\label{15-1}
\begin{split}
&\|{x}_i(t+1)-{z}_i(t)\|\\
&\leq \frac{\sqrt{n}h\kappa_2\kappa_3+\kappa_1}{\mu}\eta \\
&~~~+\frac{1}{\mu}\eta \|b_i^t-\nabla_2f_i^t(x_i(t),x_i(t))\|\\
&~~~+\frac{\sqrt{n}\kappa_2}{\mu}\eta \|c_i^t-\nabla g_i^t(x_i(t))\|\\
  \end{split}
\end{eqnarray}
Denote $d_i(t)=x_i(t+1)-z_i(t)$, there holds
\begin{eqnarray*}\begin{split}
x_i(t+1)=\sum_{j\in\mathcal{N}_i(t)}a_{ij}(t)x_j(t)+d_i(t).
\end{split}\end{eqnarray*}
It can be rewritten as
\begin{eqnarray*}\begin{split}
x_i(t)&=\sum_{j\in\mathcal{N}_i(t)}a_{ij}(t-1)x_j(t-1)+d_i(t-1)\\
&=\Big([A(t-1)]_{i\cdot}\otimes I_m\Big)x(t-1)+d_i(t-1)\\
&=\Big([\Phi(t,0)]_{i\cdot}\otimes I_m\Big)x(0)\\
&~~~~+\sum_{s=1}^t\Big([\Phi(t,s)]_{i\cdot}\otimes I_m\Big)d(s-1)\\
\end{split}\end{eqnarray*}
where ${x}(t)=col\{x_i(t)\}_{i\in\mathcal{V}}$ and ${d}(t)=col\{d_i(t)\}_{i\in\mathcal{V}}$. By the fact that $\Phi(t,s)$ is a doubly
stochastic matrix, one has
\begin{eqnarray*}\begin{split}
\bar{x}(t)=\Big(\frac{1}{n}1_n^T\otimes I_m\Big)x(0)+\sum_{s=1}^t\Big(\frac{1}{n}1_n^T\otimes I_m\Big)d(s-1).
\end{split}\end{eqnarray*}
Combining above two qualities yields that for any $i\in\mathcal{V}$,
\begin{eqnarray*}\begin{split}
&\|x_i(t)-\bar{x}(t)\|\\
&\leq\left\|\left([{\Phi}(t,0)]_{i\cdot}-\frac{1}{n}{1}_n^T\right)\otimes I_m\right\|\|x(0)\|\\
&~~~~+\sum_{s=1}^t\left\|\left([{\Phi}(t,s)]_{i\cdot}-\frac{1}{n}{1}_n^T\right)\otimes I_m\right\|\|d(s-1)\|.
\end{split}\end{eqnarray*}
Using Lemma \ref{le1}, $\|x(0)\|\leq n\kappa$ and (\ref{15-1}) yields that
\begin{eqnarray*}\begin{split}
&\|x_i(t)-\bar{x}(t)\|\\
&\leq \mathcal{C}n^{\frac{3}{2}}\sqrt{m}\kappa\lambda^t+\frac{\mathcal{C}{n}^{2}\sqrt{m}h\kappa_2\kappa_3+\mathcal{C}n^{\frac{3}{2}}\sqrt{m}\kappa_1}{\mu}\eta\sum_{s=1}^t\lambda^{t-s}\\
&~~~ +\frac{\mathcal{C}\sqrt{nm}}{\mu}\eta\sum_{s=1}^t\lambda^{t-s} \\
&~~~\sum_{i=1}^n\|b_i^{s-1}-\nabla_2f_i^{s-1}(x_i(s-1),x_i(s-1))\|\\
&~~~+\frac{\mathcal{C}n\sqrt{m}\kappa_2}{\mu}\eta\sum_{s=1}^t\lambda^{t-s}\sum_{i=1}^n\|c_i^{s-1}-\nabla g_i^{s-1}(x_i({s-1}))\|.
\end{split}\end{eqnarray*}
Using the inequality $(a+b+c+d)^2\leq 4(a^2+b^2+c^2+d^2)$ for any $a,b,c,d\in\mathbb{R}$ yields
\begin{eqnarray}\label{15-2}\begin{split}
&\|x_i(t)-\bar{x}(t)\|^2\\
&\leq 4\mathcal{C}^2n^3m\kappa^2\lambda^t+\frac{4(\mathcal{C}{n}^{2}\sqrt{m}h\kappa_2\kappa_3+\mathcal{C}n^{\frac{3}{2}}\sqrt{m}\kappa_1)^2}{\mu^2(1-\lambda)^2}\eta^2\\
&~~~+\frac{4\mathcal{C}^2{nm}}{\mu^2}\eta^2\Big(\sum_{s=1}^t\lambda^{t-s} \\
&~~~\sum_{i=1}^n\|b_i^{s-1}-\nabla_2f_i^{s-1}(x_i(s-1),x_i(s-1))\|\Big)^2\\
&~~~+\frac{4\mathcal{C}^2n^2{m}\kappa_2^2}{\mu^2}\eta^2\Big(\sum_{s=1}^t\lambda^{t-s}\\
&~~~\sum_{i=1}^n\|c_i^{s-1}-\nabla g_i^{s-1}(x_i(s-1))\|\Big)^2.
\end{split}\end{eqnarray}
Using Cauchy-Schwarz inequality, we have
\begin{eqnarray}\label{15-4}\begin{split}
&\Big(\sum_{s=1}^t\lambda^{t-s} \sum_{i=1}^n\|b_i^{s-1}-\nabla_2f_i^{s-1}(x_i(s-1),x_i(s-1))\|\Big)^2\\
&\leq n\sum_{s=1}^t\lambda^{t-s}\sum_{s=1}^t\lambda^{t-s}\\
&~~~\sum_{i=1}^n\|b_i^{s-1}-\nabla_2f_i^{s-1}(x_i(s-1),x_i(s-1))\|^2\\
&\leq \frac{n}{\lambda(1-\lambda)}\sum_{s=0}^t\lambda^{t-s}\sum_{i=1}^n\|b_i^s-\nabla_2f_i^s(x_i(s),x_i(s))\|^2.
\end{split}\end{eqnarray}
We can also have
\begin{eqnarray}\label{15-5}\begin{split}
&\left(\sum_{s=1}^t\lambda^{t-s}\sum_{i=1}^n\|c_i^{s-1}-\nabla g_i^{s-1}(x_i(s-1))\|\right)^2\\
&\leq n\sum_{s=1}^t\lambda^{t-s}\sum_{s=1}^t\lambda^{t-s}\\
&~~~\sum_{i=1}^n\|c_i^{s-1}-\nabla g_i^{s-1}(x_i(s-1))\|^2\\
&\leq \frac{n}{\lambda(1-\lambda)}\sum_{s=0}^t\lambda^{t-s}\sum_{i=1}^n\|c_i^s-\nabla g_i^s(x_i(s))\|^2.
\end{split}\end{eqnarray}
Substituting (\ref{15-4}) and (\ref{15-5}) into (\ref{15-2}) and taking the summation with respect to $t=0, \cdots, T$ result  in that
\begin{eqnarray}\label{15-8}\begin{split}
&\sum_{t=0}^T\|x_i(t)-\bar{x}(t)\|^2\\
&\leq \frac{4\mathcal{C}^2n^3m\kappa^2}{1-\lambda}+\frac{4(\mathcal{C}{n}^{2}\sqrt{m}h\kappa_2\kappa_3+\mathcal{C}n^{\frac{3}{2}}\sqrt{m}\kappa_1)^2}{\mu^2(1-\lambda)^2}\\
&~~~(T+1)\eta^2+\frac{4\mathcal{C}^2{n^2m}}{\mu^2\lambda(1-\lambda)^2}\eta^2\\
&~~~\sum_{t=0}^T\sum_{i=1}^n\|b_i^t-\nabla_2f_i^t(x_i(t),x_i(t))\|^2\\
&~~~+\frac{4\mathcal{C}^2n^3{m}\kappa_2^2}{\mu^2\lambda(1-\lambda)^2}\eta^2\sum_{t=0}^T\sum_{i=1}^n\|c_i^t-\nabla g_i^t(x_i(t))\|^2.
\end{split}\end{eqnarray}
By the proof of Lemma \ref{le9-1}, we can conclude that for any $\nu_1,\nu_2\in(0,1)$, there holds
\begin{eqnarray}\label{15-9}\begin{split}
&\sum_{t=0}^T\eta^2\sum_{i=1}^n\|b_i^t-\nabla_2f_i^t(x_i(t),x_i(t))\|^2\\
&\leq n\sigma_1^2(T+1)\eta^2+\sigma_1^2\ln{\frac{1}{\nu_1}}
\end{split}\end{eqnarray}
with  probability at least $1-\nu_1$, and
\begin{eqnarray}\label{15-10}\begin{split}
&\sum_{t=0}^T\eta^2\sum_{i=1}^n\|c_i^t-\nabla g_i^t(x_i(t))\|^2\\
&\leq n\sigma_2^2(T+1)\eta^2+\sigma_2^2\ln{\frac{1}{\nu_2}}
\end{split}\end{eqnarray}
with  probability at least $1-\nu_2$. Substituting (\ref{15-9}) and (\ref{15-10}) into (\ref{15-8}) gives that  with  probability at least $1-(\nu_1+\nu_2)$,
\begin{eqnarray}\label{15-11}\begin{split}
&\sum_{t=0}^T\|x_i(t)-\bar{x}(t)\|^2\\
&\leq \frac{4\mathcal{C}^2n^3m\kappa^2}{1-\lambda}+\frac{4(\mathcal{C}{n}^{2}\sqrt{m}h\kappa_2\kappa_3+\mathcal{C}n^{\frac{3}{2}}\sqrt{m}\kappa_1)^2}{\mu^2(1-\lambda)^2}\\
&~~~(T+1)\eta^2+\frac{4\mathcal{C}^2{n^3m}\sigma_1^2}{\mu^2\lambda(1-\lambda)^2}(T+1)\eta^2\\
&~~~+ \frac{4\mathcal{C}^2{n^2m}\sigma_1^2}{\mu^2\lambda(1-\lambda)^2}\ln{\frac{1}{\nu_1}}+\frac{4\mathcal{C}^2n^3{m}\sigma_2^2\kappa_2^2}{\mu^2\lambda(1-\lambda)^2}\ln{\frac{1}{\nu_2}}\\
&~~~+\frac{4\mathcal{C}^2n^4{m}\sigma_2^2\kappa_2^2}{\mu^2\lambda(1-\lambda)^2}(T+1)\eta^2.
\end{split}\end{eqnarray}
Letting $\nu_1=\nu_2=\frac{\nu}{2}$ immediately implies (\ref{15}).
By (\ref{eq11-2}), taking the square on both sides of (\ref{15-1}) and using the same methods as (\ref{15-8})-(\ref{15-11}), one can obtain (\ref{15*}).

\emph{H.  Proof of Theorem 2}

In Lemma \ref{le2}, letting $\hat{x}={x}_i(t+1)$, ${y}={z}_i(t)$, $w=x^*_t$ and ${s}=\eta b_i^t+\eta c_i^t{y}_i(t)$,
with reference to the proof of Lemma \ref{le5}, one can easily obtain that
  \begin{eqnarray}\label{b3}
\begin{split}
&-\zeta f^t(x_i(t), x^*_t)\\
&\leq \sum_{i=1}^n\eta \langle b_i^t-\nabla_2 f_i^t(x_i(t),{x}_i(t)), {x}^*_t-{x}_i(t)\rangle\\
&~~~+\sum_{i=1}^n\eta \langle  c_i^t{y}_i(t)-\nabla g_i^t(x_i(t)){y}_i(t), {x}^*_t-{x}_i(t)\rangle\\
\end{split}\end{eqnarray}
\begin{eqnarray*}\begin{split}
&~~~+\sum_{i=1}^n\eta^2\|b_i^t-\nabla_2 f_i^t(x_i(t),{x}_i(t)) \|^2\\
&~~~+\sum_{i=1}^n\eta^2 \|c_i^t{y}_i(t)-\nabla g_i^t(x_i(t)){y}_i(t)\|^2\\
&~~~+2\sum_{i=1}^n\|x_i(t)-x_i(t+1)\|^2+n\rho^2\eta^2\\
&~~~-\eta \sum_{i=1}^n(y_i(t))^Tg_i(x_i(t))+\kappa_2\eta \sum_{i=1}^n\|y_i(t)-\bar{y}(t)\|\\
&~~~+\sum_{i=1}^n\Big(\mathcal{D}_\phi(x^*_t, x_i(t))-\mathcal{D}_\phi(x^*_{t+1}, x_i(t+1))\Big)\\
&~~~+\|x_i(t)-\bar{x}(t)\|^2+2n^2L^2\eta^2 +n\ell\|x^*_{t+1}-x^*_{t}\|.
  \end{split}
\end{eqnarray*}
 Note that (\ref{28}) also holds if $\eta_t=\eta$, letting $\eta_t=\eta$ in (\ref{28}), substituting the achieved inequality into (\ref{b3}) and  summing over $t=0, \cdots, T$, we have
\begin{eqnarray}\label{b4}\begin{split}
&-\sum_{t=0}^T \zeta f^t(x_i(t), x^*_t)\\
&+\sum_{t=0}^T\sum_{i=1}^n\eta \Big(g_i^t(x_i(t))\Big)^Ty-\Big(1+(T+1)\eta \Big)\frac{n\|y\|^2}{2}\\
&\leq \sum_{t=0}^T\sum_{i=1}^n\eta \langle b_i^t-\nabla_2 f_i^t(x_i(t),{x}_i(t)), {x}^*_t-{x}_i(t)\rangle\\
&~~~+\sum_{t=0}^T\sum_{i=1}^n\eta \langle c_i^t{y}_i(t)-g_i^t(x_i(t)){y}_i(t), {x}^*_t-{x}_i(t)\rangle\\
&~~~+\sum_{t=0}^T\sum_{i=1}^n\eta^2\|b_i^t-\nabla_2 f_i^t(x_i(t),{x}_i(t)) \|^2\\
&~~~+\sum_{t=0}^T\sum_{i=1}^n\eta^2 \|c_i^t{y}_i(t)-\nabla g_i^t(x_i(t)){y}_i(t)\|^2\\
&~~~+(3\kappa_2+2\sqrt{n}\kappa_2)\sum_{t=0}^T\sum_{i=1}^n\eta \|y_i(t)-\bar{y}(t)\|+4nK\kappa^2\\
&~~~+(n\kappa_2^2+n^2\kappa_2^2)(T+1)\eta^2+\frac{9}{2}\sum_{t=0}^T\sum_{i=1}^n\|y_i(t)-\bar{y}(t)\|^2\\
&~~~+2\sum_{t=0}^T\sum_{i=1}^n\|x_i(t)-x_i(t+1)\|^2+n\rho^2(T+1)\eta^2\\
&~~~+\sum_{t=0}^T\|x_i(t)-\bar{x}(t)\|^2+2n^2L^2(T+1)\eta^2 +n\ell\Theta_T.
\end{split}\end{eqnarray}
By the proof of Lemma \ref{le9-1}, we know that both the fourth term and the fifth term on the right-hand side of (\ref{b4}) have a high probability bound $\mathcal{O}\Big(T\eta^2+\ln{\frac{1}{\nu}}\Big)$, together with (iv), (v) in Lemma \ref{le6}, (\ref{13}), (\ref{13*}) in Lemma \ref{le9-1}, and (\ref{15}), (\ref{15*}) in Lemma \ref{le10},  we can achieve that with probability at least $1-\nu$,
\begin{eqnarray}\label{b4-1}\begin{split}
&-\sum_{t=0}^T \zeta f^t(x_i(t), x^*_t)\\
&+\sum_{t=0}^T\sum_{i=1}^n\eta \Big(g_i^t(x_i(t))\Big)^Ty-\Big(1+ (T+1)\eta \Big)\frac{n\|y\|^2}{2}\\
&\leq\mathcal{O}\Big(T\eta^2+\Theta_T+\ln\frac{1}{\nu}\Big).
\end{split}\end{eqnarray}
Letting $y=0$ and dividing by $\zeta$ yield  (\ref{7-5}). Moreover, letting $y=\frac{\big[\sum_{t=0}^T\sum_{i=1}^n\eta g_i^t(x_i(t))\big]_+}{n(1+(T+1)\eta )}$ results in that
\begin{eqnarray*}\begin{split}
&\Big(\sum_{t=0}^T\sum_{i=1}^n\eta g_i^t(x_i(t))\Big)^Ty-\Big(1+(T+1)\eta \Big)\frac{n\|y\|^2}{2}\\
&=\frac{\eta^2\big[\sum_{t=0}^T\sum_{i=1}^n g_i^t(x_i(t))\big]_+^2}{2n(1+(T+1)\eta)}.
\end{split}\end{eqnarray*}
Together with (\ref{32}), (\ref{7-4}) and (\ref{b4-1}), it immediately implies (\ref{7-6}). The proof of Theorem \ref{TH2} is completed.

\end{document}